\documentclass[11pt,a4paper]{article}

\usepackage{lmodern} 
\usepackage[utf8]{inputenc} 
\usepackage[T1]{fontenc} 
\usepackage{microtype} 
\usepackage[english]{babel}

\usepackage{color,graphicx}
\usepackage[percent]{overpic}

\usepackage{amsmath,dsfont,url,amssymb}
\usepackage{slashed,cancel}
\usepackage[usenames,dvipsnames]{xcolor}
\usepackage[colorlinks=true, linkcolor=black, citecolor=ForestGreen]{hyperref}
\usepackage{mdframed}

\def \d {\partial}

\renewcommand{\Im}[0]{\operatorname{Im}}

\usepackage{amssymb}

\def\ben{\begin{equation}}
\def\een{\end{equation}}
\def\half{{\textstyle{\frac{1}{2}}}}

   \let\d=\delta \let\e=\varepsilon

\let\w=\omega \let\G=\Gamma

\def\be{\begin{equation}}
\def\ee{\end{equation}}
\def\beq{\begin{equation}}
\def\eeq{\end{equation}}
\def\ba{\begin{array}}
\def\ea{\end{array}}

\def\dalemb#1#2{{\vbox{\hrule height .#2pt
       \hbox{\vrule width.#2pt height#1pt \kern#1pt
               \vrule width.#2pt}
       \hrule height.#2pt}}}

\newcommand{\bea}{\begin{eqnarray}}
\newcommand{\eea}{\end{eqnarray}}

\makeatletter
\newcommand*\bigcdot{\mathpalette\bigcdot@{.5}}
\newcommand*\bigcdot@[2]{\mathbin{\vcenter{\hbox{\scalebox{#2}{$\m@th#1\bullet$}}}}}
\makeatother

\renewcommand{\eqref}[1]{(\ref{#1})}

\def\Im{{{\frak{Im}}}}

\def\ocal{{\mathcal{O}}}

\newcommand{\E}{\mathcal{E}}

\renewcommand{\Im}[0]{\operatorname{Im}}

\begin{document}
\frenchspacing

\title{Thermalization, Viscosity and
\\ the Averaged Null Energy Condition}
\author{Luca V. Delacr\'etaz$^\flat$, Thomas Hartman$^\sharp$, Sean A. Hartnoll$^\flat$ and Aitor Lewkowycz$^\flat$ \\ {\it $^\flat$ Department of Physics, Stanford University,}\\
{\it Stanford, California, USA} \\
{\it $^\sharp$ Department of Physics, Cornell University,} \\ {\it Ithaca, New York, USA}
}

\date{}
\maketitle


\begin{abstract}
    
We explore the implications of the averaged null energy condition for thermal states of relativistic quantum field theories. A key property of such thermal states is the thermalization length. This lengthscale generalizes the notion of a mean free path beyond weak coupling,
and allows finite size regions to independently thermalize. Using the eigenstate thermalization hypothesis, we show that thermal fluctuations in finite size `fireballs' can produce states that violate the averaged null energy condition if the thermalization length is too short or if the shear viscosity is too large. These bounds become very weak with a large number $N$ of degrees of freedom but can constrain real-world systems, such as the quark-gluon plasma. 
    
\end{abstract}

\section{Introduction: ANEC and ETH}

Kovtun, Son, and Starinets conjectured a lower bound on shear viscosity, suggesting that fundamental principles of quantum statistical mechanics could usefully constrain the dynamics of strongly interacting, non-quasiparticle systems \cite{Kovtun:2004de}.  The recent proof of a bound on chaos realizes this intuition to an extent \cite{Maldacena:2015waa}, although Lyapunov growth is not directly related to physical observables such as transport coefficients. Transport is instead controlled by local thermalization. A lower bound on the viscosity has yet to be established, and the status of such a bound remains inconclusive \cite{Cremonini:2011iq}, but the shear viscosity was recently found to have an {\it upper bound} set by the thermalization timescale \cite{Hartman:2017hhp}. That bound follows from requiring diffusion to be causal, in the spirit of earlier observations \cite{Baier:2007ix}.

The use of fundamental constraints on quantum field theories (QFTs) to bound observables is at the core of the `bootstrap' approach to conformal field theories \cite{Poland2016}. As part of this endeavour, the
averaged null energy condition (ANEC) has been proven in unitary, Lorentz invariant QFTs \cite{Klinkhammer:1991ki,Faulkner:2016mzt, Hartman:2016lgu}.
The ANEC constrains the extent to which energy density can be negative in any state $\psi$ of a quantum field theory:
\begin{equation}\label{eq:ANEC}
  \langle \E_+ \rangle_{\psi} \equiv \int dx^+ \langle T_{++} \rangle_{\psi} \geq 0 \,.
\end{equation}
Here $T_{++}$ is a null component of the energy-momentum tensor, and the integral is over a null ray. Historically, this integrated energy condition was found to be sufficient to prove several theorems in general relativity \cite{Roman:1986tp,Borde:1987qr}. More recently, applied to conformal field theories, the ANEC has been shown to constrain stress tensor couplings \cite{Hofman:2008ar}, current-energy couplings \cite{Hofman:2009ug,Chowdhury:2012km}, other operator product expansion coefficients \cite{Cordova:2017zej}, and operator dimensions \cite{Cordova:2017dhq}. The derivation of the ANEC from the monotonicity of relative entropy in \cite{Faulkner:2016mzt} also demonstrates a direct connection between the ANEC and the real-time dynamics of quantum information. 

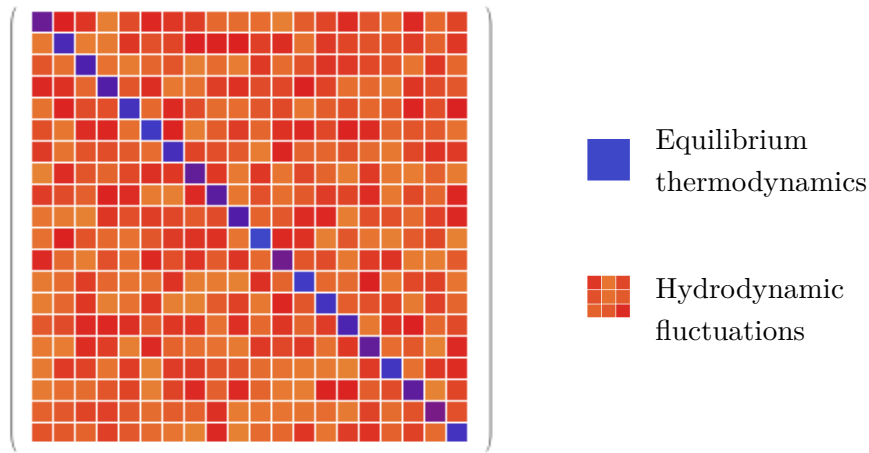
\begin{figure}[!b]
\hspace{2cm}
\begin{overpic}[width=400px,tics=10]{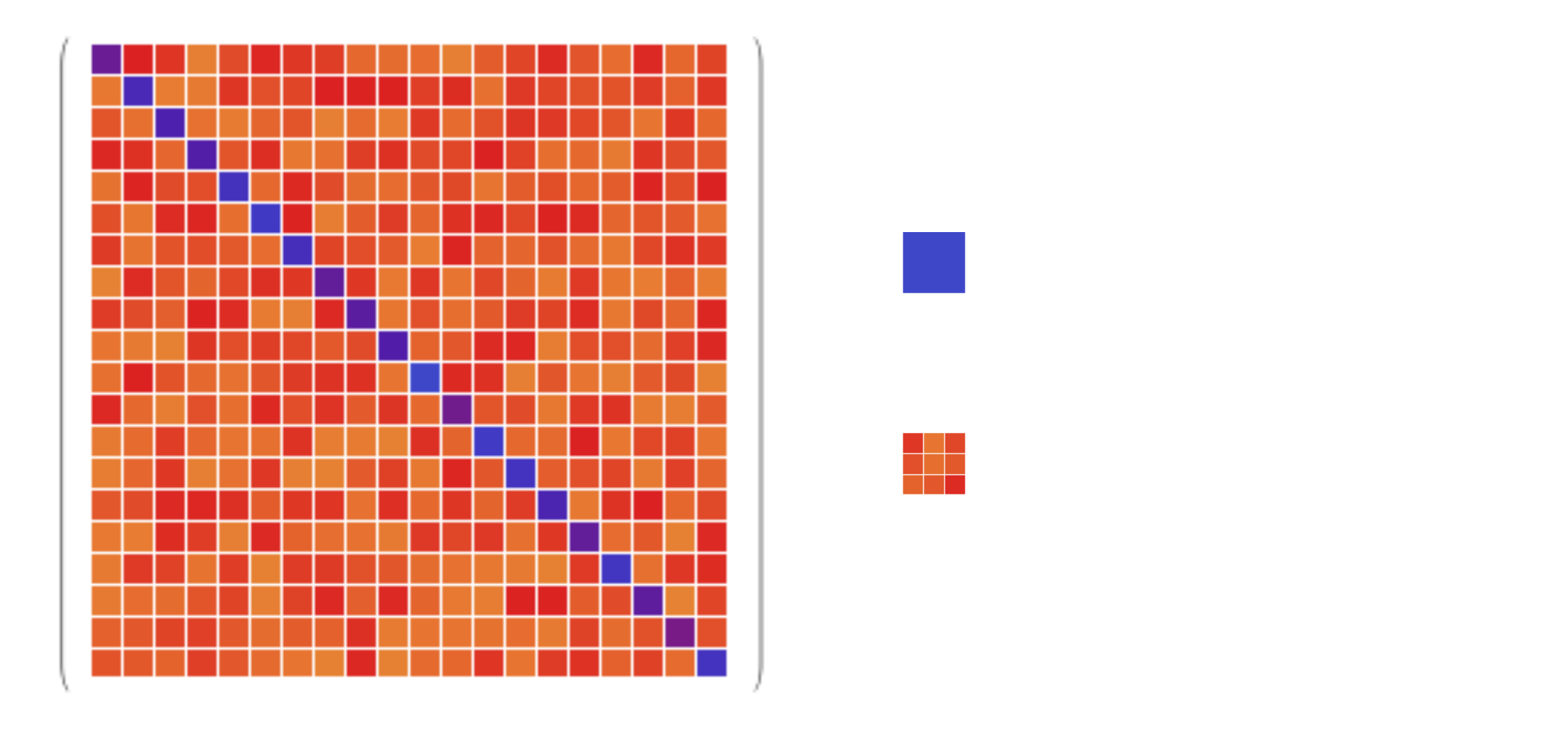}
 \put (64,31) {Equilibrium }
 \put (64,27) {thermodynamics}
 \put (64,17) {Hydrodynamic}
 \put (64,13) {fluctuations} 
\end{overpic}    
    \caption{\small Structure of the matrix $\langle a|T_{++}|b\rangle$, in eigenstates with energy near some fixed $E$. The diagonal elements are related to thermal expectation values $\langle T_{++}\rangle_T$, while the exponentially suppressed, off-diagonal elements are related to the correlation functions $\langle T_{++}T_{++}\rangle_T$. The ANEC constrains the magnitude of the off-diagonal elements in terms of those along the diagonal. }
    \label{fig:matrix}
\end{figure}

In this work explore the extent to which the ANEC constrains nonzero temperature transport physics. To apply the ANEC to transport, we consider the matrix elements of the operator $\E_+$ in highly excited pure states. See figure \ref{fig:matrix}.  In a basis of eigenstates, the diagonal elements of this matrix are set by equilibrium thermodynamics. The off-diagonal elements are related to energy-momentum fluctuations. We will
see that at long wavelengths these control the hydrodynamic relaxation toward equilibrium. The ANEC requires this matrix to be positive semi-definite, otherwise it would be possible to construct linear combinations of eigenstates that violate the ANEC. Therefore, the off-diagonal elements (hydrodynamics) will be constrained by the diagonal elements (equilibrium thermodynamics).  This is a nonzero temperature version of the interference effect described in \cite{Cordova:2017zej}.

In order to regulate the infinite null integral in the ANEC (\ref{eq:ANEC}) it is essential to consider a finite size, locally thermalized `fireball' state. The smaller the region, the stronger the thermal fluctuations. So long as the region is larger than the thermalization length, it can remain in local thermal equilibrium. We will find that if the thermalization length becomes too short or if the viscosity becomes too large, then thermal fluctuations allow a  superposition of microstates to violate the ANEC. However, over most but not all of our parameter space, saturation of the bound on thermalization length roughly coincides with the limit of validity of our computation due to various finite size effects. We will describe these below.

To obtain quantitative constraints on hydrodynamics, we will employ the eigenstate thermalization hypothesis (ETH). 
This is the expectation that highly excited energy eigenstates in a quantum system are effectively thermal.
The ETH determines the form of the matrix elements of a local operator $\ocal(x)$ between two highly excited energy eigenstates $| a \rangle$ and $| b \rangle$ \cite{PhysRevA.43.2046,PhysRevE.50.888,rigoleth,eth}. In a translationally invariant system it is important to keep track of the conserved momentum as well as the energy of these states \cite{eth}. We denote the 4-momentum of the state $|a\rangle$ by $p_a = (E_a,\vec p_a)$ and write 
\be\label{tppe}
\langle a| \ocal(0) |b\rangle = \langle \ocal(0) \rangle_{T} \delta_{ab} + t(p_a, p_b)R_{ab} \,.
\ee
The first term on the right hand side is the expectation value of $\ocal$ in thermal equilibrium with temperature $T$ corresponding to the energy $E_a$. The states $|a\rangle$ are taken to be within an energy window about some large reference energy $E$ such that, in the large volume limit, the temperature is the same for all the states considered. The second term describes thermal fluctuations, with the independent random variables $R_{ab}$ having zero mean and unit variance. The magnitude of the smooth function $t$ and the distribution of the $R_{ab}$ is to be determined by the requirement that the ansatz (\ref{tppe}) recovers results for a thermal state. The individual fluctuation terms will be exponentially suppressed in the thermal entropy with respect to the diagonal contributions. However, the random matrix $R_{ab}$ exhibits strong eigenvalue repulsion that can overcome the suppression of the individual entries. Our use of the ETH will not require knowledge of the probability distribution from which the $R_{ab}$ are drawn, beyond their independence.

Inserting $T_{++}$ into the ETH ansatz (\ref{tppe}), and being careful to regularize the total volume, will allow us to bound hydrodynamic fluctuations by thermal expectation values, as outlined above. The explicit form of the hydrodynamic fluctuations --- contained in $t(p_a,p_b)$ --- will be consistent with conservation of the total energy and momentum, so that the non-randomness of these special quantities is properly accounted for.

Our final result bounds a complicated function of transport coefficients and thermalization length and time. For moderate values of these quantities, the bound is schematically
\be\label{eq:roughbound}
s\ell_{\rm th}^3 \gtrsim 1 \ .
\ee
Here $s$ is the entropy density and $\ell_\text{th}$ is the thermalization length. It is therefore a lower bound on the total entropy in a thermal volume. From a microscopic perspective, equation (\ref{eq:roughbound}) is tautological to a degree (that said, we have not seen this simple point made elsewhere). It expresses the fact that in order for a region of extent $\ell_\text{th}$ to in fact be thermal, it must contain sufficiently many degrees of freedom. Furthermore, in order to use the ETH a sufficiently large number of degrees of freedom are needed, precisely along the lines of (\ref{eq:roughbound}).\footnote{However, the $s$ in (\ref{eq:roughbound}) arises in our computation as a thermal expectation value, rather than directly as the size of the ETH random matrix. The latter quantity in fact cancels out of our final answer.} This means that in regimes where (\ref{eq:roughbound}) is violated, our derivation of the bound is likely invalid. Such regions of parameter space remain excluded, but for the thermodynamic reason just described rather than the ANEC. Nonetheless, our result in (\ref{eq_bound_manip}) below is more fine-grained than (\ref{eq:roughbound}), with a nontrivial function of the viscosities, sound speed and thermalization length and time appearing on the right hand side. In particular, in the limits of small sound speed or large viscosities, the right hand side will be seen to become parametrically large, establishing bounds from the ANEC within the regime of validity of the computation. Furthermore, if certain finite size boundary effects are small --- a fact we cannot ascertain in the current approach --- then for order one values of the parameters the bound is more accurately $s \ell_\text{th}^3 \gtrsim 500$. See the left hand plot in figure \ref{fig:exclusion} below. This may be stronger than needed for local thermalization, and hence a nontrivial consequence of the ANEC.

Hydrodynamics is the effective theory of long wavelength excitations of a thermal medium, and holds below a momentum cutoff $\Lambda \equiv \ell_{\rm th}^{-1}$. The result (\ref{eq:roughbound}) is therefore an upper bound on the cutoff, $\Lambda \lesssim T s_o^{1/3}$. Here $s_o \equiv s/T^3$ is a dimensionless measure of degrees of freedom. In large $N$ theories, $s_o$ is large and hence the bound becomes weak. We proceed to apply our bound to the strongly interacting quark-gluon plasma. Combined with a previous upper bound on the shear viscosity, $\eta/s \lesssim T \ell_\text{th}$ \cite{Hartman:2017hhp}, the quark-gluon plasma lies close to a `kink' in an exclusion plot in the plane of allowed values of $\eta/s$ and $T \ell_\text{th}$. See figure \ref{fig:expt} below.

\section{Hydrodynamic contribution to fluctuations}
\label{sec:ttt}

The function $t(p_a,p_b)$ in the ETH (\ref{tppe}) can be directly related to the momentum-space thermal Wightman function as follows. For the state $|a\rangle$ to describe thermal equilibrium we must have
\be
\langle a| \ocal(x) \ocal(0) |a\rangle = \langle \ocal(x) \ocal(0) \rangle_{T} \,.
\ee
We will take $|a\rangle$ to be at rest in the `lab frame' so that the spatial components $\vec p_a = 0$. Now insert a complete set of energy eigenstates on the left hand side of this relation, and Fourier transform both sides. This gives
\be\label{eq:Gsum}
G_{\ocal\ocal}(p) = \sum_b \delta^4(p + p_a - p_b)| \langle a|\ocal(0)|b\rangle|^2 \,.
\ee
Here $G_{\ocal\ocal}(p)$ is the momentum-space thermal Wightman function at temperature $T$. We next evaluate the matrix elements using the ETH ansatz (\ref{tppe}). The local operator $\ocal(0)$ cannot change the energy density of the state $|a\rangle$ and therefore the only $|b\rangle$ states that contribute to the sum in (\ref{eq:Gsum}) are within the energy window to which the ETH can be applied. We will be interested in a nonzero external 4-momentum $p$ and hence the diagonal term in (\ref{tppe}) does not contribute. The fluctuation terms give
\bea
G_{\ocal\ocal}(p) & = & \sum_b \delta^4(p + p_a - p_b) |t(p_a, p_b)|^2 |R_{ab}|^2 \\
& = & \sum_b  \delta^4(p + p_a - p_b) |t(p_a, p_b)|^2 \,. \label{gbf}
\eea
In the second line we have performed the average over the $R_{ab}$ variables. Finally, to invert (\ref{gbf}), introduce the density of states $\Omega(p_b)$, dependent on both energy and momentum, such that
\be
\sum_b \to \int dp_b \Omega(p_b) \,.
\ee
Inverting \eqref{gbf}  gives
\be\label{tgvv}
|t(p_a, p_b)|^2 = \frac{G_{\ocal \ocal}(p_b - p_a)}{  \Omega(p_b)} \,.
\ee
See  \cite{Brehm:2018ipf,Romero-Bermudez:2018dim,Hikida:2018khg} for recent discussions of this formula in $1+1$ dimensional conformal field theories.

The momentum space Wightman function can be related to the thermal retarded Green's function $G^R_{\ocal \ocal}$ using the fluctuation-dissipation theorem:
\be
G_{\ocal \ocal}(p)  = \frac{2 \, \text{Im} G^R_{\ocal \ocal}(p)}{\displaystyle 1 - e^{-p^0/T}} \,.
\ee
The imaginary part of the retarded Green's function is a spectral function that can be directly related to the on-shell excitations in the system, as we see shortly. Using the above expression we can write (\ref{tgvv}) as 
\be\label{eq:tfluc}
|t(p_a, p_b)|^2 = \frac{2 \, \text{Im} G^R_{\ocal \ocal}(p_b - p_a)}{\Omega(p_b) - \Omega(p_a)} \,.
\ee
Here we used the fact that for small excitations about equilibrium, $p^\mu = (E+\omega,\vec{k})$,
\be\label{eq:Om}
\Omega(p) = \Omega(E) \, e^{\omega /T} \, .
\ee
This follows from $\Omega \sim e^S$, with $S$ the entropy, together with $\Delta S = \Delta E \frac{\partial S}{\partial E}=\Delta E/T$. The expression \eqref{eq:Om} is exact at large volume so long as $\omega,\vec{k}$ are not extensive (i.e. correspond to a vanishing energy and moment density at large volume). In general, a shift $\Delta \vec P$ in the total spatial momentum also leads to an additional change in entropy $\Delta S = \vec v \cdot \Delta \vec P/T$, with $\vec v$ the velocity. However, because we are considering thermal and near-thermal states with non-extensive momentum, then $\vec v$ itself will be zero at large volume. Therefore this shift in the entropy is subleading compared to the $\Delta E/T$ shift.

The imaginary part of the retarded Green's function is an important physical quantity that determines the rate of heating if the system is driven by a source for the operator $\ocal$. In particular, in hydrodynamic regimes where the system is probed at low energies and long wavelengths, the retarded Green's functions for conserved densities and their currents can be determined systematically. Specifically, the hydrodynamic regime is
\be
\omega \lesssim \frac{2 \pi}{\tau_\text{th}} \,, \quad k \lesssim \frac{2\pi}{\ell_\text{th}} \,. 
\ee
The thermalization time $\tau_\text{th}$ and length $\ell_\text{th}$ will be discussed in more detail below. In Appendix \ref{app_hyd} we obtain the hydrodynamic limit of $G^R_{T_{++}T_{++}}$ in a general relativistic quantum field theory. We assume a thermal state with zero density of charge for any internal global symmetries of the theory, otherwise the hydrodynamic modes are more complicated. To be precise about notation, the light-cone coordinates and momenta are defined in terms of the Minkowski variables $x^\mu = (t,\vec x)$ and $p^\mu = (p^0,k_x, k_\perp)$ as
\begin{equation}\label{eq:null}
    x^{\pm} = t\pm x\, , \qquad
    k_{\pm} = \frac{1}{2}(-p^0 \pm k_x)\, , \qquad
    T_{++} = \frac{1}{4}T_{tt}  + \frac{1}{2} T_{tx} + \frac{1}{4} T_{xx} \, ,
\end{equation}
with the transverse momenta $k_\perp     = (k_y,k_z)$. In order to connect with the ANEC operator (\ref{eq:ANEC}), we restrict attention to the $++,++$ components of the Green's function in the hydrodynamic limit. The result from Appendix \ref{app_hyd} is then:
\begin{equation}\label{eq:hydromain}
\begin{split}
  \rho(p) \equiv  \mbox{Im}\, G^R_{T_{++}T_{++}}(p)\ = \frac{F(p)}{\omega^2 + D_\perp^2 k^4} + \frac{G(p)}{\left(\w^2 - c_s^2 k^2\right)^2 + \G_s^2 \w^2 k^4} \, ,
\end{split}
\end{equation}
where $k = |\vec k|$.
Here we have shown explicitly the singular behavior due to the diffusive and sound modes. These are respectively \cite{Son:2007vk}
\be\label{eq:diff}
\omega = - i D_\perp k^2 \,, \qquad D_\perp = \frac{\eta}{\e + P} \,,
\ee
where $\e + P$ is the enthalpy density ($\e$ is the energy density and $P$ the pressure) and $\eta$ is the shear viscosity, and
\be\label{eq:sound}
\omega = \pm c_s k - i \frac{\Gamma_s}{2} k^2 + \cdots \,, \qquad \Gamma_s = \frac{\zeta + \frac{4}{3} \eta}{\e+ P} \,.
\ee
Here $c_s$ is the sound speed and $\zeta$ is the bulk viscosity. Throughout we will work in 3+1 spacetime dimensions. In 2+1 dimensions hydrodynamic fluctuations can be very strong and the physics is likely more subtle \cite{Kovtun:2012rj}. In 1+1 dimensions the kinematics is markedly different, for example in a 1+1 conformal field theory there is no hydrodynamic regime \cite{Herzog:2007ij}. The functions $F$ and $G$ in (\ref{eq:hydromain}) are given in Appendix \ref{app_hyd}.

Inserting the hydrodynamic Green's function (\ref{eq:hydromain}) into the expression (\ref{eq:tfluc}) for the fluctuations in the ETH ansatz, it is clear the viscosities $\eta$ and $\zeta$ directly control the long wavelength fluctuations of the local operator $T_{++}(0)$. From (\ref{eq:null}) we see that $T_{++}$ involves the energy, momentum and pressure. Conservation of total energy and momentum implies that these quantities do not fluctuate strongly on long wavelengths. The dominant contribution to the Green's function  (\ref{eq:hydromain}) at the longest wavelengths instead comes from pressure fluctuations; the pressure is of course not a conserved quantity. This is important for our use of the ETH for the ANEC operator. However, a direct use of the ETH expression (\ref{tppe}) to evaluate the ANEC operator (\ref{eq:ANEC}) leads to long distance divergences because of the integral over a null ray in the homogeneous equilibrium state. Fortunately, the structure of thermal equilibrium itself can regulate these divergences, as we now explain.

\section{Local thermalization}

We have already introduced the thermalization length $\ell_\text{th}$. This generalizes the notion of a mean free path and is the statement that generic (non-hydrodynamic) correlation functions in a thermal state decay exponentially with spatial distance: $\langle \ocal(\vec x) \ocal(0) \rangle_\text{c} \sim e^{-|\vec x|/\ell_\text{th}}$. It follows that thermal equilibrium can be self-consistently established within a region of spatial extent $\ell_\text{th}$. Indeed, in relaxing to equilibrium, a system will first locally thermalize, so that regions of size $\ell_\text{th}$ each have their own temperatures $T_1, T_2, T_3, \ldots$ and their own velocities $v_1, v_2, v_3\, \ldots$ (in a minimal relativistic theory where the only conserved quantities are energy and momentum). At this point hydrodynamics takes over and describes the slow global equilibration of temperature and velocity. Hydrodynamics is a coarse-grained description valid on wavelengths $2 \pi/k \gtrsim \ell_\text{th}$. The hydrodynamic variables are precisely the modes that have survived local thermalization, the
local temperature and velocity fields, $T(x)$ and $v(x)$, as well as the associated thermal current and stress tensor. In short, the fact of local thermalization is essential for the very existence of hydrodynamics.

Given that thermal equilibrium can be established locally in any region $A$ of size $L \gtrsim \ell_\text{th}$, we can apply the ETH to this region directly. In particular we can regulate the long distance divergence by restricting to states that are thermalized in $A$ but in the vacuum outside. This could be an expanding thermal `fireball' or else a spatially confined thermal region (a `fireplace'). To construct such states we introduce a projection operator 
that acts as the identity inside $A$ and gradually projects the complement into the vacuum over scales of order $\ell_\text{th}$.
This smoothness ensures that the expansion of a fireball state is itself described by hydrodynamics and is hence subluminal. 
Acting on such states, the ANEC operator
\be
\E_+ \equiv \int dx^+ T_{++}(x^+, x^- = 0, x^\perp = 0) \,,
\ee
will be equivalent to the projection $P_A \E_+  P_A^\dagger$, that only acts in the fireball. Positivity of $\E_+$ implies that $P_A \E_+  P_A^\dagger$ is again a positive operator.

Local thermalization implies that for operators that only act in the region $A$, we can write a new ETH ansatz in terms of a basis of states $|i\rangle$ also supported in $A$, obtained by $|i \rangle = P_A |a\rangle$. For example, these could be eigenstates of the projected Hamiltonian $P_A H  P_A^\dagger$. There are many fewer such states than eigenstates of the full system. An operator a distance $\ell_\text{th}$ or more from the boundary of the region cannot distinguish these states from full energy eigenstates.\footnote{Local versions of ETH have also been discussed in e.g. \cite{Garrison:2015lva,PhysRevE.97.012140}. The key difference in the construction here is that the exterior region is in vacuum.} Therefore, up to edge effects, these states behave as energy and momentum eigenstates of the full system. The decoupling of spatial regions on the scale $\ell_\text{th}$ means that in the limit $L \gg \ell_\text{th}$ edge effects are parametrically suppressed. This is in contrast to e.g. the Casimir effect, where edge effects are important in order to see that the ANEC is obeyed \cite{Graham:2005cq}. Even with $L \gtrsim \ell_\text{th}$, the discussion of local thermalization above means that the ETH ansatz will hold, with eigenstates that are now weakly inhomogeneous over the scale of the fireball, and eigenvalue repulsion will again lead to the tendency for fluctuations to produce states that can violate the ANEC. We proceed to mostly neglect edge effects in the following. However, as we will discuss in more detail below equation (\ref{eq_bound_manip}), we cannot strictly exclude the (interesting) possibility that edge effects conspire to always overcome the effects of fluctuations, such that the ANEC is obeyed across all of parameter space. We will also neglect any subluminal growth of the fireball as it is traversed by the null ray, which leads to at most an order one change in the effective length $L$ of the region.

The matrix elements of the projected ANEC operator between fireball states $|i\rangle$ and $|j\rangle$ can therefore be written
\bea
\E^{A}_{+ij} \equiv \frac{1}{2L}\langle i | P_A \E_+  P_A^\dagger |j \rangle &=&  \frac{1}{2L}\int_{-L}^{L} e^{-i (p_+^i - p_+^j) x^+} dx^+ \langle i|T_{++}(0)|j\rangle\\
&=&  \langle T_{++}(0) \rangle_{T}\delta_{ij}  + \frac{ \sin\left( L(p_+^i - p_+^j)\right)}{L(p_+^i - p_+^j)}t(p_i,p_j)R_{ij}
\,. \label{eq:withd}
\eea
In the first line we have used the fact that $|i\rangle$ and $|j\rangle$ are (to an excellent approximation) eigenstates of $P_+$. The second line used the ETH ansatz for $T_{++}(0)$, restricted to the fireball states $|i\rangle$, as described above. The result of the $x^+$ integral is to set $p_+^i \approx p_+^j$, up to corrections of order $L^{-1}$.

With the explicit matrix $\E^{A}_{+ij}$ at hand, we are finally in a position to obtain constraints from the positivity of this matrix.

\section{Constraints on thermalization and transport}

A positive operator is positive in any subspace. 
That is, the matrix
\be\label{mfirst}
M_{ij} \equiv \Theta(p_i) \E^{A}_{+ij} \Theta(p_j) \,,
\ee
is positive for any projection $\Theta(p)=\Theta(p^0,\vec k)$ equal to one in some range and zero elsewhere. We will choose $\Theta$ to be supported in a window defined by
\begin{align}\label{eq:window}
    |p^0-E|  < \half \Delta\omega
    \qquad \hbox{and} \qquad 
    k_x = k_y = k_z =  \frac{\pi}{L} \,.
\end{align}
In this way, by choosing $L \gtrsim \ell_\text{th}$ and $\Delta \omega \lesssim 2\pi/\tau_\text{th}$ we restrict to excitations that are described by hydrodynamics. We have put the momenta equal to the smallest values that are possible given the finite extent $L$ of the fireball, as this choice is found to lead to the strongest bounds.
The positive matrix $M$ then has rank 
\be
\begin{split}
N 
    \equiv \int dp_i \, \Omega(p_i)\Theta(p_i) 
    &= \left(\frac{\pi}{L}\right)^3
    \int_{E-\half \Delta \omega}^{E+\half\Delta \omega}dp^0 \Omega(p^0) \\
    &= 2\Omega(E)(\pi/L)^3 T \sinh \frac{\Delta \omega}{2 T}\, ,
\end{split}
\ee
where $\int dp \equiv \int dp^0 (\pi/L)^3\sum_{\vec{k}}$, and we used the expression \eqref{eq:Om} for the density of states.

The matrix $M$ is a sum of diagonal and off-diagonal terms. From (\ref{eq:withd}), the diagonal terms are large and effectively constant within the subspace, given by $\langle T_{++}(0)\rangle_{T} = \frac{1}{4} (\varepsilon + P)$. In fact there are variations in this expectation value between the different eigenstates due to the finite volume, we will discuss these later. Positivity of the matrix therefore requires that the most negative eigenvalue of the off-diagonal part of the matrix in (\ref{eq:withd}), restricted to the subspace, have absolute value smaller than $\tfrac{1}{4}(\varepsilon + P)$. The minimal eigenvalue (of the off-diagonal part of the matrix) obeys 
\be\label{genmin}
\lambda_\text{min}^2 \geq \frac{1}{N} \sum_{i,j}\frac{ \sin^2\left( L(p_+^i - p_+^j)\right)}{L^2(p_+^i - p_+^j)^2} |t(p_i,p_j)|^2 \Theta(p_i)\Theta(p_j) \ .
\ee
This result follows from the fact that for a general symmetric matrix $A$, the eigenvalue with the largest magnitude has $\lambda^2 \geq \frac{1}{N} \sum_{i,j,k}A_{ij}A_{jk}$. Using (\ref{eq:withd}) and (\ref{mfirst}) in this expression, together with the randomness of $R_{ij}$ --- specifically the fact that the different components are independently random and have unit variance --- leads to \eqref{genmin}. We are also assuming that the eigenvalue distribution of $R_{ij}$ is symmetric, so that the largest and smallest eigenvalues have the same magnitude. A closely related expression has recently been used in \cite{Dymarsky:2017zoc}. Note that we do not need to know the higher moments of the distribution of the $R_{ij}$. These moments contain information about higher point thermal correlation functions. Because we have only used the variance of the $R_{ij}$, only the two point thermal correlation function, contained in $|t(p_i,p_j)|^2$ via (\ref{eq:tfluc}), enters our inequality.

We can now use the explicit form (\ref{eq:tfluc}) for $|t(p_i,p_j)|^2$
in the inequality (\ref{genmin}), with the null energy spectral function given by (\ref{eq:hydromain}). The constraint that $\lambda_\text{min} \leq \tfrac{1}{4}(\varepsilon + P)$ then becomes
\begin{equation}\label{eq_bound_manip}
\begin{split}
(\varepsilon+P)^2 
	&\geq \frac{32}{N} \int dp_i \int dp_j \frac{\Omega(p_i) \Omega(p_j)}{\Omega(p_i) - \Omega(p_j)} 
	\frac{\sin^2 L(p^i_+ - p^j_+)}{L^2(p^i_+ - p^j_+)^2}
	\rho(p_i-p_j)   \\
	&= 
	\frac{64\pi^3}{T L^3  \sinh \frac{\Delta \omega}{2 T}}
	\int_{-\frac{\Delta\omega}{2}}^{\frac{\Delta \omega}{2}} d\omega_i \int_{-\frac{\Delta\omega}{2}}^{\frac{\Delta \omega}{2}} d\omega_j
	\frac{\sin^2  \frac{L}{2}(\omega_i - \omega_j)   }{L^2 (\omega_i - \omega_j)^2} \frac{\rho(\omega_i - \omega_j, k_\text{min})}{e^{-\omega_j/T} - e^{-\omega_i/T}}
		\\
	&= \frac{64\pi^3}{L^3 \sinh \frac{\Delta \omega}{2T}}
	\int_{- \Delta \omega}^{\Delta \omega}d\omega \frac{ \sin^2\frac{L \omega}{2}}{L^2 \omega^2}  \frac{\sinh\frac{1}{2T}(\Delta \omega - |\omega|) }{\sinh \frac{\omega}{2T}}\rho(\omega,k_\text{min}) \,.
\end{split}
\end{equation}
In the second line we set $p^0_{i,j} = E + \omega_{i,j}$. In the last line we performed the integral over $\omega_i+\omega_j$. The $\sin^2 \frac{L\omega}{2} $ term in the final expression comes from the geometric integral in (\ref{eq:withd}) and has nothing to do with random matrices. The momenta are restricted to the single value given in (\ref{eq:window}), so that $\vec k_i - \vec k_j = 0$.  However, in the spectral density $\rho$ the momentum difference cannot be set to be strictly zero.
We are using the infinite volume Green's function in $\rho$ instead of the finite volume Green's function. The two will agree on scales $k \gtrsim k_\text{min} \equiv \pi/L$, the smallest momentum that can be resolved by the finite volume Green's function, up to boundary effects. Boundary effects in the Green's function arise on timescales such as the Thouless time, over which diffusion propagates across the entire system and hence hydrodynamic modes becomes sensitive to the edges of the fireball. We need to worry about such effects because contributions to the integral in (\ref{eq_bound_manip}) from frequencies below the inverse Thouless time will be important for our more interesting bounds. In Appendix \ref{sec:finiteV} we argue that extra terms in the Green's function due to finite size effects at most change our results by an overall order one number when $L \gg \ell_\text{th}$. The role of the Thouless time in eigenstate thermalization has recently been discussed in \cite{Dymarsky:2018ccu}.

Expectation values also differ between infinite and finite volume systems. We are aiming to make a statement involving infinite volume expectation values, but it is the finite volume expectation value that appears in our derivation. Finite volume corrections are concerning because even a small correction of order $1/L^{3}$ to the left hand side of (\ref{eq_bound_manip}) competes with the right hand side, for any size of the fireball. In Appendix \ref{sec:finiteV} we argue that such terms are not expected to dominate the fluctuation contribution, at least when the fluctuations become large. If the finite-size corrections are also parametrically enhanced when the fluctuations become large, then there is still a nontrivial bound, but the left-hand size of \eqref{eq_bound_manip} should be interpreted to include these corrections.

It is instructive to write the bound (\ref{eq_bound_manip}) in terms of dimensionless variables. Because we are considering quantum field theories at zero charge density, the thermodynamic relation $\e+P = sT$ holds, where $s$ is the entropy density. We can furthermore write $s = s_o T^3$, where $s_o$ is dimensionless (but can be temperature dependent in general). It is clear from the overall inverse powers of $L$ in (\ref{eq_bound_manip}), that the smaller we can make the size of the fireball, the stronger the bound we will obtain. However, we also require $L \gtrsim \ell_\text{th}$, both in order for the fireball to be locally thermalized and also for the longest wavelength modes inside the fireball to be described by hydrodynamics. Therefore, to get the strongest possible bound we set $L = \ell_\text{th}$. Of course, to safely neglect edge effects, as we have done in several places, one should set $L$ to be a multiple of $\ell_\text{th}$. An explicit hierarchy between $L$ and $\ell_\text{th}$ can easily be re-inserted --- as we will see shortly --- and weakens the constraints by numerical factors but does not remove them. Similarly, the bound is strongest if we integrate over the largest possible range of frequencies $\Delta \omega$ that are compatible with hydrodynamics. Therefore we set $\Delta \omega = 2\pi/\tau_\text{th}$. Putting all of this together, the inequality (\ref{eq_bound_manip}) then takes the functional form
\be\label{eq:dimensionless}
s_o \geq {\mathcal F}\left(\frac{\eta}{s}, \frac{\zeta}{s}, c_s, \ell_\text{th} T, \tau_\text{th} T \right) \,.
\ee
All of the quantities appearing in the function ${\mathcal F}$ --- to be obtained by performing the integral in (\ref{eq_bound_manip}) --- are dimensionless in units with $\hbar = c = k_B = 1$. The inequality (\ref{eq:dimensionless}) therefore amounts to a constraint on the viscosities, sound speed and thermalization time and length, given the dimensionless entropy $s_o$.

\section{Results}

\begin{figure}[h]
\centering
\includegraphics[width=200px]{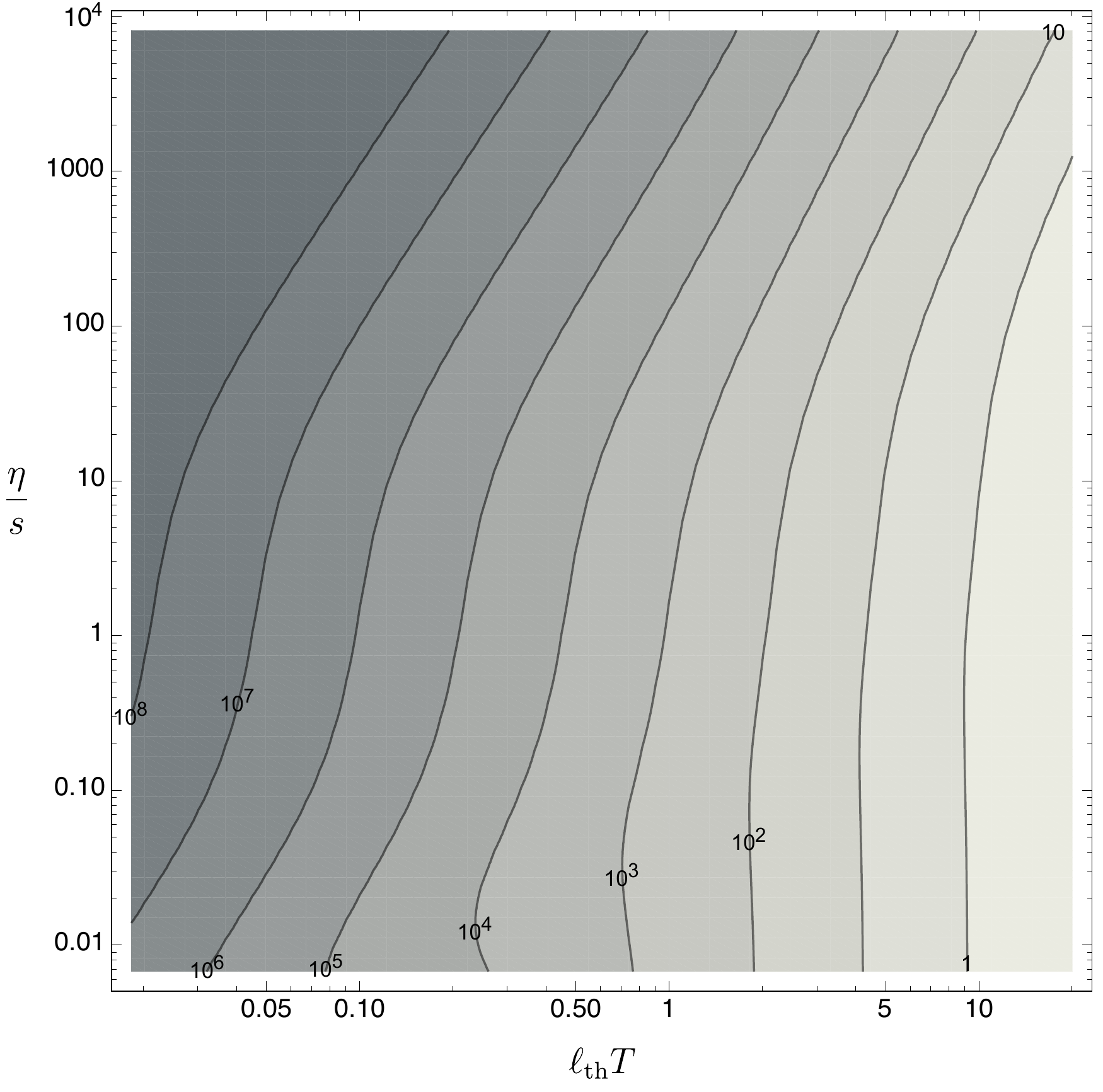}\hspace{0.1cm}
\includegraphics[width=200px]{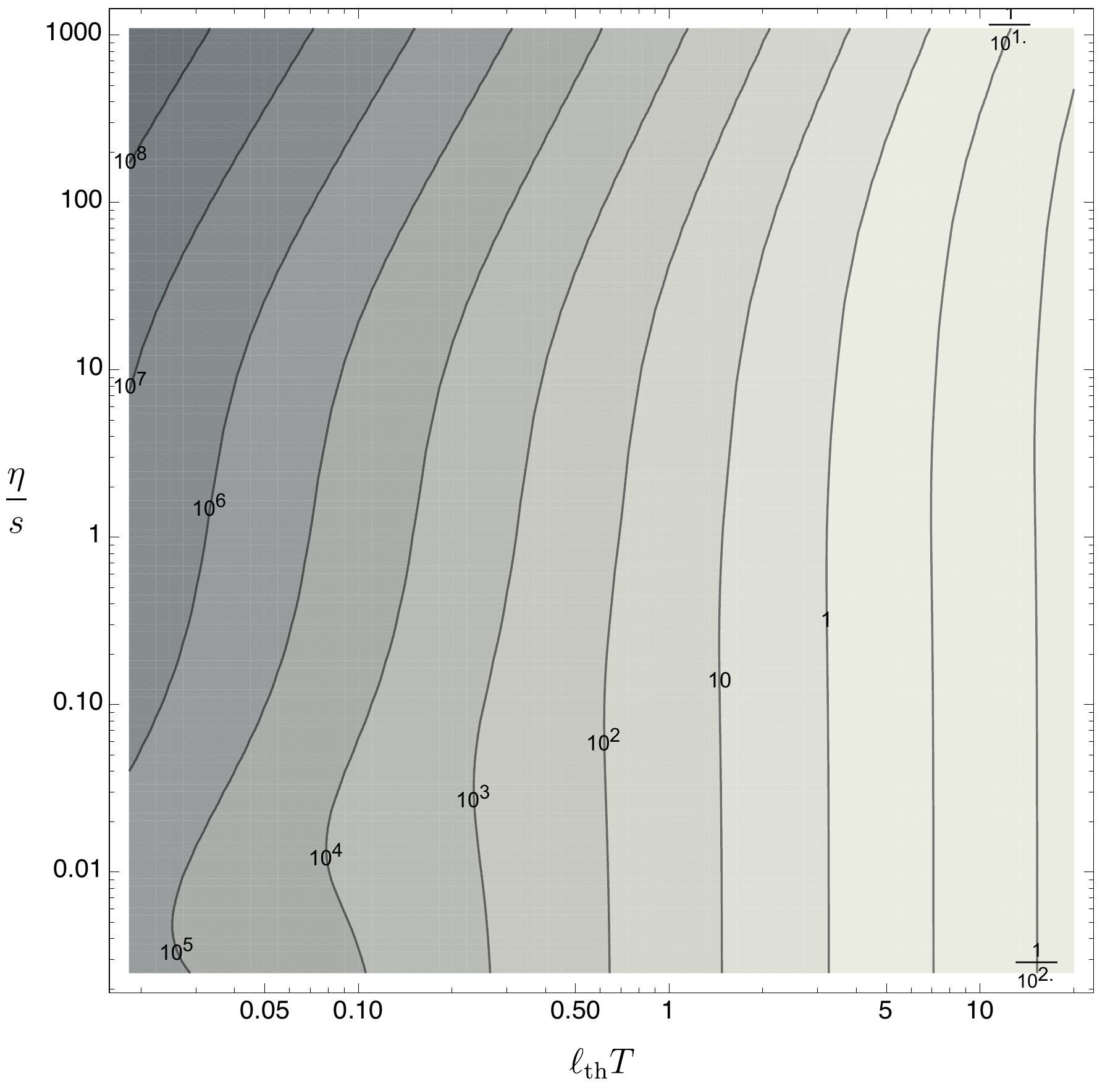}
\caption{\small Contours of constant ${\mathcal F}_\text{CFT}\left(\eta/s, \ell_\text{th} T \right)$. For a given `dimensionless entropy' $s_o$ regions with ${\mathcal F}_\text{CFT} \geq s_o$ are inconsistent with the ANEC. The plot on the left is for a fireball of size $L = \ell_\text{th}$. The plot on the right is for size $L = 3 \ell_\text{th}$. Taking a larger fireball weakens the bound numerically but does not qualitatively change the exclusion regions. \label{fig:exclusion}}
\end{figure}

To get a sense of the consequences of (\ref{eq:dimensionless}), consider first the case of a conformal field theory in which $\zeta = 0$ and $c_s^2 = 1/3$. For simplicity in this case, let us further temporarily set $\tau_\text{th} = \ell_\text{th}$. This leaves a constraint in a two dimensional parameter space $s_o \geq {\mathcal F}_\text{CFT}\left(\eta/s, \ell_\text{th} T \right)$. A contour plot showing the parameter regions excluded for different values of $s_o$ is shown in figure \ref{fig:exclusion}.
We have made two plots, one with $L = \ell_\text{th}$ and one with $L = 3 \ell_\text{th}$. A hierarchy between $L$ and $\ell_\text{th}$ is necessary in order to strictly neglect boundary effects due to the finite size of the fireball. The qualitative form of the excluded regions is seen to be the same in both cases. For some given $s_o$, the figure shows that for at any fixed $\eta/s$ there is a lower bound on $\ell_\text{th} T$, and that at any fixed $\ell_\text{th} T$, there is a (somewhat weak) upper bound on $\eta/s$. We now proceed to understand these main features in figure \ref{fig:exclusion} analytically, working with general --- not necessarily conformal --- theories. The function ${\mathcal F}$ in (\ref{eq:dimensionless}) simplifies in several important limits. 

\subsection*{Short thermalization lengths}

In the limit of small $\ell_\text{th}T$ and $\tau_\text{th} T$ --- with other quantities held fixed --- then ${\mathcal F} \sim 1/\ell_\text{th}^3$, with no singular dependence on $\tau_\text{th}$. This limit is responsible for the lower bound on $\ell_\text{th} T$ visible in figure \ref{fig:exclusion}. Simplifying the final integral in (\ref{eq_bound_manip}), in this limit the bound becomes
\be\label{eq:small}
\left. S \right|_{\ell_\text{th}} \equiv s \ell_\text{th}^3 \geq 4 \pi^3 \int_0^\infty \frac{y (c_1 + c_2 y^2)}{(e^y-1)(c_3 + c_4 y^2)} dy \,.
\ee
Here the left hand side is the total entropy in a region of size $\ell_\text{th}^3$. This is indeed the natural quantity that controls the effects of thermal fluctuations. The constants
\begin{align}
c_1 =  \left(\frac{\zeta}{s} + \frac{4}{3} \left(1 + 2 c_s^2 + \frac{4}{3} c_s^4 \right) \frac{\eta}{s} \right)\,,  \quad & c_3 =  c_s^4 \,, \\
c_2  = \frac{16}{9} \frac{\eta}{s} \left( \frac{\zeta}{s} + \frac{1}{3}\frac{\eta}{s} \right) \left(  \frac{\zeta}{s} + \frac{4}{3} \frac{\eta}{s} \right)\,, \quad & c_4  = \left(  \frac{\zeta}{s} + \frac{4}{3} \frac{\eta}{s} \right)^2\,.
\end{align}
The most important corollary of (\ref{eq:small}) is that $\ell_\text{th}$ is necessarily lower bounded for any fixed values of the viscosities, sound speed and entropy density. The bound becomes weak in large $N$ theories, where thermodynamic fluctuations are suppressed. For example, if $s \sim N^2$ then $\ell_\text{th} \gtrsim N^{-2/3}$. Finally, in the `non-relativistic limit' $c_s \ll 1$, this bound simplifies to $s \ell_\text{th}^3 \geq \pi^4/(4 c_s^2)$. In this limit the bound becomes parametrically strong --- recall the discussion below (\ref{eq:roughbound}) above.

\subsection*{Large viscosities}

The expression (\ref{eq:small}) does not correctly capture the limits of small or large viscosities. These limits on the viscosity do not commute with the small $\ell_\text{th}$ limit.
At large viscosities, $\eta/s \gg 1$ and $\zeta/s \gg 1$, with other quantites held fixed, the bound becomes
\be\label{eq:etaup}
s \ell_\text{th}^3 \geq \frac{128 \pi^3}{9}  \frac{\eta}{s} \frac{\zeta + \frac{1}{3} \eta}{\zeta + \frac{4}{3} \eta} f(\ell_\text{th}T,\tau_\text{th} T) \,,
\ee
where the function
\be\label{eq:fun}
f(\ell_\text{th}T,\tau_\text{th} T) = \int_0^{\pi/(\tau_\text{th} T)} \frac{d x}{x} \frac{\sin^2 \left(\ell_\text{th} T x \right)}{(\ell_\text{th} T)^2} \left(\coth x -  \coth \frac{\pi}{\tau_\text{th} T} \right) \,.
\ee
Because the right hand side of (\ref{eq:etaup}) grows at large $\eta/s$, it amounts to an upper bound on $\eta/s$ at fixed thermalization length and time. Correspondingly, the lower bound on $s \ell_\text{th}^3$ becomes parametrically strong in the limit where $\eta/s \gg 1$, again see the discussion below (\ref{eq:roughbound}) above.

Taking $\ell_{\rm th} \sim \tau_{\rm th}$, the function in (\ref{eq:fun}) scales as $f \sim 1/(\ell_\text{th} T)$ at large $\ell_\text{th} T$ and goes to a constant at small $\ell_\text{th} T$. In particular, at large $\ell_\text{th} T$ the bound becomes $\eta/s \lesssim 10^{-2} s_o (\ell_{\rm th} T)^4$, which determines the asymptotic behavior of the contours in the left hand plot of figure \ref{fig:exclusion}.
In this limit, the upper bound (\ref{eq:etaup}) on $\eta/s$ is therefore weaker than the upper bound $\eta/s \lesssim \tau_\text{th} T$ obtained from causality in \cite{Hartman:2017hhp}, assuming that $\ell_\text{th} \sim \tau_\text{th}$. This latter bound saturates in weakly coupled theories, which is furthermore the circumstance under which $\eta/s$ is expected to be large. However, if $\ell_\text{th} T$ is small, then the bound (\ref{eq:etaup}) is stronger for $1/(\ell_\text{th} T)^3 \ll s_o \ll 1/(\ell_\text{th} T)^2$. We will include both bounds in our discussion of the quark-gluon plasma below.

\subsection*{Small viscosities}

The bound (\ref{eq:small}) becomes weak at small viscosities, while in fact the lower bound on $\ell_\text{th}$ survives in the limit $\eta = \zeta = 0$. The only contribution to the integral (\ref{eq_bound_manip}) in this limit comes from the diffusive and sound poles in the spectral density. With vanishing viscosities, the integral can be done exactly, and the bound takes the form
\be\label{eq:van}
s \ell_\text{th}^3 \geq \frac{16 \pi^4}{3} + \frac{c_5 }{\ell_\text{th} T} \left(\coth\frac{\sqrt{3} \pi c_s}{2 \ell_\text{th} T} - \coth \frac{\pi}{\tau_\text{th} T} \right) \,.
\ee
The term in brackets should be set to zero if it becomes negative. This occurs when the sound pole is above the energy cutoff $2 \pi/\tau_\text{th}$ when evaluated at the momentum cutoff $\pi/\ell_\text{th}$. Here the coefficient
\be\label{eq:c5}
c_5 = \frac{4(3+c_s^2)(1+3 c_s^2) \pi^3 \sin^2 \frac{\sqrt{3} c_s \pi}{2}}{3 \sqrt{3} c_s^3} \,.
\ee
Of course the limit of vanishing viscosity is unlikely to be physical. The point is that a lower bound on the thermalization length survives in this limit. This fact establishes that the lower bound on thermalization length does not become small for any value of the viscosity. Furthermore, the fact that the bound remains finite rather than diverging as the shear viscosity becomes small (with, for example, the bulk viscosity set to zero) shows that we do not obtain an absolute lower bound on the shear viscosity.

The limit of small viscosities, however, has an additional subtlety: If the bare viscosity becomes too small, the classical hydrodynamic expansion breaks down \cite{Kovtun:2011np}. We now address this point.

\subsection*{Loop corrections within hydrodynamics}

We have assumed throughout that the hydrodynamic Green's function (\ref{eq:hydromain}) holds for all frequencies and wavevectors $\omega \lesssim 2\pi/\tau_\text{th}$ and $k \lesssim \pi/\ell_\text{th}$. While it is always true that only the hydrodynamic modes are present over these scales, their dynamics becomes strongly coupled if the viscosity becomes small. In such regimes, hydrodynamic loop corrections are important and the Green's function no longer takes the `classical' hydrodynamic form (\ref{eq:hydromain}). In 3+1 dimensions, the most dangerous nonanalytic correction is a `late time tail' that invalidates the hydrodynamic expansion beyond first order, so that in the $k=0$ Green's function \cite{Kovtun:2011np}
\be
\eta \to \eta + 0.012 (i-1)\omega^{1/2} \frac{(s T)^{3/2}}{\eta^{3/2}} + \cdots \,.
\ee
The nonanalytic correction is large for small viscosities.
In order to neglect this term --- and hence for the classical hydrodynamic Green's function to remain valid --- we must restrict to frequencies
\be
\omega \lesssim \Lambda \approx 7000 \, s_o^2 T \left(\frac{\eta}{s}\right)^5 \,. \ee
This additional constraint means that our answer will change in the region of parameter space where the new cutoff is stronger than our previous one, i.e. where $\Lambda < 2\pi/\tau_\text{th}$. That is, for $1/(\tau_\text{th} T) \lesssim 10^3 s_o^2 (\eta/s)^5$ we cannot trust our answer. While this formally excludes both the interesting limits of  parametrically fast thermalization and parametrically small viscosities, in fact, because of the large numerical factor of order $10^3 s_o^2$, it will have no impact in our discussion of the quark-gluon plasma below (where $s_o \approx 20$). Specifically, this region of parameter space is entirely outside of the exclusion plot shown in figure \ref{fig:expt} below.

Loop corrections also renormalize the viscosity. We are working with the physical, renormalized viscosity throughout. While the leading order renormalization tends to push the viscosity away from zero \cite{Kovtun:2011np}, the effects of high order loops are not known.

\subsection*{Large thermalization lengths}

At large thermalization lengths and times, i.e. taking $\ell_\text{th} T \sim \tau_\text{th} T \gg 1$ with other quantities held fixed, the dominant contribution to the integral (\ref{eq_bound_manip}) is again from the diffusive and sound poles in the spectral density. We obtain
\be\label{eq:long}
s \ell_\text{th}^3 \geq \frac{16 \pi^4}{3} + \frac{c_5}{\pi c_s} \left(\frac{2}{\sqrt{3}} - c_s \frac{\tau_\text{th}}{\ell_\text{th}} \right) \,.
\ee
The coefficient $c_5$ was defined in (\ref{eq:c5}). As in (\ref{eq:van}), the term in brackets should be set to zero when it is negative. In fact, (\ref{eq:long}) is just the small viscosity result (\ref{eq:van}), additionally expanded for large thermalization length and time. This is because the viscosities appear in our bound through combinations such as $\eta k_\text{min}^2 \sim \eta/\ell_\text{th}^2$, so that small viscosity behaves similarly to large thermalization length. Weakly coupled theories with a large thermalization length would, in fact, typically be expected to have a large viscosity $\eta \sim \ell_\text{th}$.

Taking (\ref{eq:long}) together with the result (\ref{eq:small}) for small thermalization lengths we see that, for fixed values of the viscosities and sound speed, the bound indeed takes the schematic form $s \ell_\text{th}^3 \gtrsim 1$ for all values of the thermalization length. In the discussion below (\ref{eq:roughbound}) we noted the large numerical factor of $16 \pi^4/3 \approx 520$. This factor may help to make this bound stronger than the statement that local thermalization requires sufficiently many degrees of freedom in a region of size the thermalization length.

\subsection*{The quark-gluon plasma}

\begin{figure}[ht]
\centering
\includegraphics[width=240px]{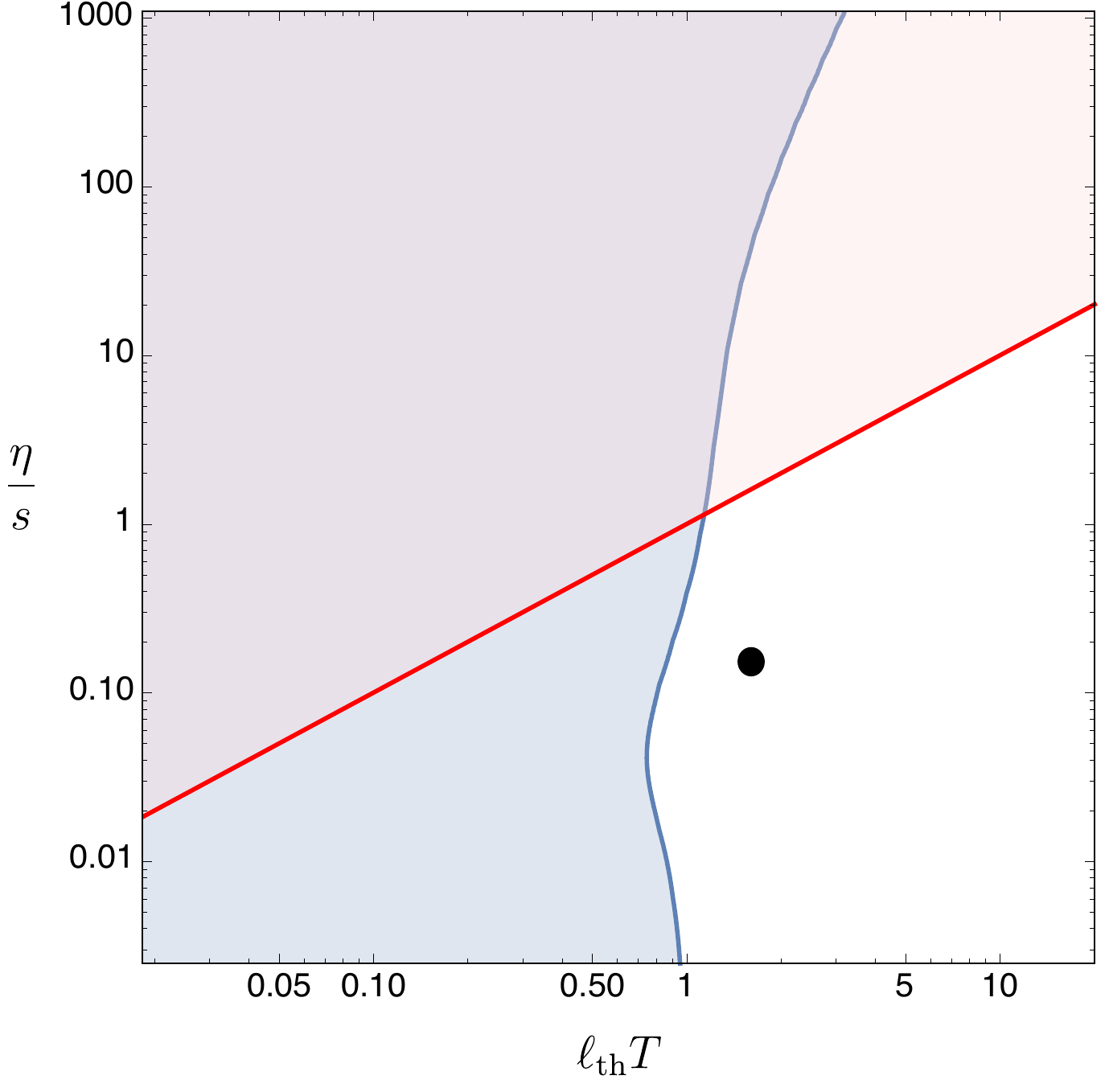}
\caption{\small Exclusion plot in the $\eta/s$ and $\ell_\text{th} T$ plane for the strongly interacting quark-gluon plasma. The blue exclusion region is from our result (\ref{eq_bound_manip}), where we have set $L = 3 \ell_\text{th}$. See main text for values of the thermodynamic and transport quantities used. The red exclusion region is from the upper bound on viscosity obtained in \cite{Hartman:2017hhp}. The black point shows representative experimental values of $\eta/s$ and $\ell_\text{th} T$.
    \label{fig:expt}}
\end{figure}

It is instructive to compare our bounds to the experimental values measured for the strongly interacting quark-gluon plasma. Some characteristic values of the relevant parameters for the quark-gluon plasma are $T \sim 330 \, \text{MeV}$, $\tau_\text{th} \sim 1 \, \text{fm/c}$, $s \sim 85 \, \text{fm}^{-3}$ \cite{Heinz:2001xi} and $\eta/s \sim 0.15 \, \hbar/k_B$ \cite{SONG2013114c,Song:2011hk}. Assuming the thermalization time is related to the thermalization length by a factor of the speed of light, this means that $\ell_\text{th} \approx 1 \, \text{fm}$. It follows that our dimensionless quantities $s_o \sim 20$ and $\ell_\text{th} T \sim 1.6$, as well as the value of $\eta/s \sim 0.15$ quoted above. In figure \ref{fig:expt} we have shown these experimental values together with the corresponding exclusion region, taking $L = 3 \ell_\text{th}$ for illustrative purposes. We furthermore used the values $c_s = 1/\sqrt{3}$ and $\zeta = 0$ --- the strongly coupled quark-gluon plasma is approximately conformal \cite{Cheng:2007jq,Borsanyi:2010cj} and has correspondingly small bulk viscosity \cite{Meyer:2007dy}. 
In figure \ref{fig:expt} we have also included the upper bound on viscosity, $\eta/s \lesssim \tau_\text{th} T$, obtained in \cite{Hartman:2017hhp}. We see that the strongly coupled quark-gluon plasma lies fairly close to a `kink' in the exclusion plot, arising from the intersection of the two different constraints. While all of the quantities shown have uncertainties in numerical factors, they are not expected to change by orders of magnitude and therefore the proximity of the data to the kink in the exclusion plot is robust.

\section{Final comments}

\subsection*{Challenges}

The essential physical content of our argument is the following. Take a thermalized but finite volume region. As the volume is decreased, thermal fluctuations become more important. If transport coefficients are too large, or if the thermalization length is too small, then fluctuations of the pressure can become large enough --- even while local thermal equilibrium is maintained --- that a certain superposition of microstates violates the ANEC.

This argument is subtle for at least two reasons. Firstly, if the fluctuations get too large then even local thermal equilibrium will not be established. This is the point discussed around (\ref{eq:roughbound}) above. Secondly, in order for fluctuations to be described by hydrodynamics, and also in order to neglect edge effects, the extent of the region cannot get too close to the thermalization length. If the region is forced to become too big, e.g. if $L \gg \ell_\text{th}$, then the bounds become very weak. For both these reasons, our argument entails a delicate balancing act that in some regimes depends on favorable numerical factors; as soon as the bounds become nontrivial they are at risk of becoming invalid. For this reason it is clearly of great interest to develop alternative approaches to this physics. We will end with some comments in this direction.

We can first make two comments regarding the presence of edge effects and the breakdown of a hydrodynamic description of the Green's function when $L \sim \ell_\text{th}$. In fact, our inequality (\ref{eq_bound_manip}) holds for all frequencies and momenta, so long as the full Green's function is used for the spectral density $\rho$, rather than just the hydrodynamic limit of the Green's function. Furthermore, in Appendix \ref{sec:finiteV} we have outlined how the entire argument can be phrased in a finite volume system from the start. This allows edge effects to be incorporated systematically. Using properties of the full finite volume Green's function, it may be possible to determine the numerical factors associated to edge effects. Alternatively, it is possible that a more local energy condition --- such as the QNEC \cite{Bousso:2015mna, Bousso:2015wca, Balakrishnan:2017bjg} --- could provide sharper constraints.

\subsection*{Conformal bootstrap}

In a conformal field theory (CFT), the conformal bootstrap can be used to constrain correlation functions using only unitarity, symmetries, and locality \cite{Poland2016}. The ANEC is a subtle consequence of unitarity plus Lorentz invariance, so ultimately, our bound follows from these same fundamental ingredients, together with the ETH ansatz. This suggests that in the CFT case, it may be possible to reproduce our bounds using the bootstrap, and perhaps to fix the numerical coefficients in the inequality. 

There are some interesting parallels between the two approaches. One route to connect them, at least in principle, is to formulate the bootstrap for the four-point functions $\langle T_{\mu\nu}T_{\alpha\beta}O_i O_j\rangle$, where $O_{i}$ and $O_j$ are other local operators. By the state-operator correspondence, this is identical to $\langle i| T_{\mu\nu}T_{\alpha\beta} |j\rangle$. Therefore, if the dimensions of $O_i$ and $O_j$ are large enough, then this correlator must encode all of hydrodynamics. High-dimension operators have been used to model thermal states in holographic 2d CFTs in e.g. \cite{Fitzpatrick:2014vua,Asplund:2014coa,Turiaci:2016cvo,Faulkner:2017hll}.

In bootstrap language, the natural setup is to fix certain contributions to the correlator, and then place upper bounds on the scale of `new physics' controlled by higher dimension operators. When $O_i$ and $O_j$ have small dimensions, this is the standard question addressed by the numerical bootstrap.  Our bound has a similar flavor, but applied to correlators with high-dimension external operators $O_i$ and $O_j$. To sharpen the comparison, in a small-$N$ CFT, we can restate our bound, combined with the diffusion bound in \cite{Hartman:2017hhp}, as
\be\label{maxlam}
\Lambda \lesssim \min\left( \frac{1}{D_\perp},\  T \right) \ .
\ee
The quantities on the right are the scales associated to hydrodynamics and thermal equilibrium, and these place an upper bound on the scale $\Lambda = \ell_{\rm th}^{-1}$ where corrections to hydrodynamics become important. As a function of the diffusion scale, \eqref{maxlam} has an intriguing similarity to the `kinks' that appear in the conformal bootstrap \cite{ElShowk:2012ht}. In principle, it should be possible to interpolate between existing bootstrap results (such as \cite{Dymarsky:2017yzx}) and \eqref{maxlam} by increasing the dimensions of the external operators, but in practice this limit is very difficult to implement numerically. Another approach to nonzero temperature bootstrap is discussed in \cite{Iliesiu:2018fao}.

The physical origin of numerical bootstrap constraints is still mysterious. On the other hand, both of the bounds in \eqref{maxlam} come from causality of the 4-point function, in two different kinematic limits. To describe these limits, insert the stress tensors symmetrically about the origin, 
\be
\langle i| T_{\mu\nu}(x^+, x^-) T_{\alpha\beta}(-x^+, -x^-)|j\rangle \ .
\ee
The diffusion bound $\Lambda \lesssim \frac{1}{D_\perp}$ comes from causality in the regime $x^+ \sim x^- \sim \ell_{\rm th}$ \cite{Hartman:2017hhp}. The ANEC was derived from causality in a different double limit, $x^+ \ll \ell_{\rm th}^2/x^- \ll \ell_{\rm th}$ \cite{Hartman:2016lgu}, so this is the limit responsible for $\Lambda \lesssim T$.

\subsection*{Nonrelativistic systems}

Many strongly interacting thermal media are not Lorentz invariant. These include electronic condensed matter systems and trapped ultracold atomic gasses. To connect with these systems our results should first be generalized to a nonzero charge or number density. The nonrelativistic and low temperature limit then corresponds to taking $\epsilon + P \to n m c^2$, where $n$ is the charge or number density, $m$ is the effective mass and we have temporarily re-inserted the speed of light $c$. Bounds on transport and thermalization have the potential to shed light on the widespread unconventional behavior observed in strongly correlated media, see e.g. \cite{Hartman:2017hhp}. It is possible that constraints on thermalization and transport similar to those we have found above will survive in the non-relativistic limit. For these constraints to be nontrivial, the speed of light would of course have to drop out upon taking the limit.

Independently of the ANEC, we noted that the bound (\ref{eq:roughbound}) follows from the fact that thermalization in a region of size $\ell_\text{th}$ requires sufficiently many degrees of freedom in that region. That statement holds for nonrelativistic theories also. And indeed, the `thermalization bound' (\ref{eq:roughbound}) directly leads to a lower bound on quasiparticle (i.e. weakly interacting) transport. The thermalization length is set by the quasiparticle mean free path in this case, assuming that the dominant quasiparticle scattering is inelastic. In such cases we can therefore express the diffusivity of any conserved density in terms of the thermalization length and the quasiparticle velocity $v$. Combining with (\ref{eq:roughbound}) we then obtain
\be
D \sim v \ell_\text{th}  \gtrsim \frac{v}{s^{1/d}} \,. 
\ee
We trivially generalized to $d$ spatial dimensions. For example, for degenerate fermions at $T < T_F$ the entropy density $s \sim k_F^{d}\, T/T_F$. Here $T_F$ is the Fermi temperature and $k_F$ the Fermi wavevector. In particular, writing the resistivity in terms of the charge diffusivity and charge compressibility
\be\label{eq:rho}
\rho = \frac{1}{\chi D} \lesssim \frac{h}{e^2} \, k_F^{2-d} \left(\frac{T}{T_F}\right)^{1/d} \,.
\ee
At $T < T_F$, this is stronger than the Mott-Ioffe-Regel bound on the resistivity (see for example the discussion in \cite{Hartnoll:2014lpa}). Note, however, that the need for the dominant scattering to be inelastic excludes disorder scattering, which would violate (\ref{eq:rho}) at low temperatures.

\vspace{1cm}

\textbf{Acknowledgments}
We would like to acknowledge several early illuminating discussions with D.~Hofman. We thank N.~Afkhami-Jeddi, A.~Dymarsky, T.~Faulkner, J.~Kaplan, R.~Mahajan and S.~McBride for useful discussions. The work of TH is supported by DOE early career award DE-SC0014123. SAH and LVD are partially supported by DOE award DE-SC0018134. AL acknowledges support from the Simons Foundation through the It from
Qubit collaboration. AL would also like to thank the Department of Physics and Astronomy
at the University of Pennsylvania for hospitality during the development of this work.

\appendix



\section{Hydrodynamic derivation of the Greens' functions}\label{app_hyd}

Relativistic hydrodynamics controls correlations functions of the stress tensor in the low momentum limit. Although only the longitudinal parts of the currents $T_{\mu i}$ are controlled by the slow dynamics of the densities $T_{\mu 0}$ via Ward identities, Lorentz invariance is sufficiently constraining to fix all components of the Green's function. These can be found by coupling the fluid to an external non-dynamical metric $g_{\mu\nu}$. In this appendix only, we work in $d$ spatial dimensions. In the Landau frame, the constitutive relation up to first order in the gradient expansion is \cite{Kovtun:2012rj}
\begin{equation}\label{eq_consti}
\langle T^{\mu\nu}\rangle
	= \epsilon u^\mu u^\nu + P \Delta^{\mu\nu} 
	- \eta \Delta^{\mu\alpha}\Delta^{\nu\beta} \left(\nabla_\alpha u_\beta + \nabla_\beta u_\alpha - \frac{2}{d} g_{\alpha\beta} \nabla_\lambda u^\lambda \right) - \zeta \Delta^{\mu\nu} \nabla_\lambda u^\lambda + \mathcal{O} (\nabla^2)\, ,
\end{equation}
where $u^\mu$ is the fluid velocity 4-vector normalized as $u^\mu u_\mu=-1$, and $\Delta^{\mu\nu} \equiv g^{\mu\nu} + u^\mu u^\nu$. In order to study linear response, \eqref{eq_consti} must be linearized around thermal equilibrium
\begin{equation}
g_{\mu\nu} = \eta_{\mu\nu} + \delta h_{\mu\nu}\, , \qquad
\epsilon = \epsilon_0 + \delta \epsilon \, , \qquad
P = P_0 + c_s^2 \delta \epsilon \, , \qquad
u^\mu = \frac{\delta^\mu_0 + \delta^\mu_i \delta v^i}{\sqrt{-g_{00}}}\, , 
\end{equation}
where the speed of sound is $c_s^2 =\partial P/\partial \epsilon$. The conservation equations $\nabla_\mu T^{\mu\nu} = 0$ are then solved for the hydrodynamic variables $\delta\epsilon,\, \delta v_i$ in terms of the sources $\delta h_{\mu\nu}$. Plugging the solutions back into the constitutive relation \eqref{eq_consti} finally gives the retarded Green's functions
\begin{equation}\label{eq_GRTT}
G^R_{T_{\mu\nu}T_{\alpha\beta}}(\omega,k)
	= - 2 \frac{\delta \langle T^{\mu\nu}\rangle}{\delta h_{\alpha\beta}}\, .
\end{equation}
The spatial components are found to be
\begin{equation}\label{eq_GRTT_res1}
\begin{split}
G^R_{T_{ij}T_{kl}}(\omega,k)
	&= \frac{\omega^2}{k^2} A_{\parallel}(\omega,k) K_{ij}K_{kl}+ \frac{\omega^2}{k^2} A_\perp(\omega,k) \left(\left(K_{ik}P_{jl} + (i \leftrightarrow  j)\vphantom{A^A_A}\right) + (k \leftrightarrow  l\vphantom{\frac{}{}})\right)\\
	&+ B_1(\omega,k) \left(\vphantom{A^A_A} K_{ij}P_{kl}+K_{kl}P_{ij}\right)  + B_2(\omega,k) P_{ij}P_{kl} \\
	&+C(\omega,k) \left(P_{ik}P_{jl}+P_{il}P_{jk}-\frac{2}{d-1}P_{ij}P_{kl}\right)\, ,
\end{split}
\end{equation}
where we defined the projectors $K_{ij} = k_i k_j/k^2$ and $P_{ij} = \delta_{ij} - K_{ij}$, and with
\begin{subequations}\label{eq_GRTT_res2}
\begin{align}
A_\parallel(\omega,k)
	&= (\epsilon+P)\frac{\omega^2}{c_s^2 k^2  - \omega^2 -i\Gamma_s k^2 \omega} \, , \\
A_\perp(\omega,k)
	&= (\epsilon+P)\frac{D_\perp k^2}{D_\perp k^2 - i\omega} \, ,\\
B_1(\omega,k)
	&= (\epsilon+P)\frac{\omega^2(c_s^2 + i\omega (2D_\perp - \Gamma_s)) }{c_s^2 k^2 - \omega^2 - i\Gamma_sk^2\omega} \, , \\
B_2(\omega,k)
	&= (\epsilon+P)\frac{i\omega (2D_\perp k^2 - i\omega)(c_s^2 - i\omega(\Gamma_s - 2D_\perp)) }{c_s^2 k^2 - \omega^2 - i\Gamma_sk^2\omega}  + \frac{2}{d-1} i\eta \omega \, ,\\
C(\omega,k)
	&= i\eta \omega \, .
\end{align}
\end{subequations}
The diffusion constant and sound attenuation rate are given by $D_\perp = \eta/(\epsilon+P)$ and $\Gamma_s=\left(\zeta + \frac{2(d-1)}{d}\eta\right)/(\epsilon+P)$ respectively. These expressions are correct up to real contact terms which do not enter in the spectral densities $\Im G^R_{T_{\mu\nu}T_{\alpha\beta }}(\omega,k)$ of interest. Other components of the Green's function \eqref{eq_GRTT} can be obtained from \eqref{eq_GRTT_res1} with Ward identities. For example one has, up to real contact terms,
\begin{equation}
G^R_{T_{0i}T_{0j}}(\omega,k) = \frac{k^kk^l}{\omega^2} G^R_{T_{ki}T_{lj}}(\omega,k)
	= K_{ij} A_\parallel(\omega,k) + P_{ij} A_\perp(\omega,k)\, .
\end{equation}
The response function that is directly related to correlations of the ANEC operator is
\begin{equation}
G^R_{T_{++}T_{++}}(p)
	= a^\mu a^\nu a^\alpha a^\beta G^R_{T_{\mu\nu}T_{\alpha\beta}}(p) \, ,
\end{equation}
with
\begin{equation}
a^\mu = 
\frac{\partial x^\mu}{\partial x^+}
= \frac{1}{2} (1,1,0,0) \, .
\end{equation}
It can be computed using \eqref{eq_GRTT_res1} and Ward identities, and is given by
\begin{equation}
\begin{split}
4G^R_{T_{++}T_{++}}
	=&\ 
	\frac{(k_x^2-k_x\omega + k_\perp^2)^4}{4\omega^2 k^6}A_\parallel
	+ \frac{k_\perp^2(k_x^2 - k_x\omega + k_\perp^2)^2}{k^6}A_\bot 
	+ \frac{k_\perp^2(k_x^2 - k_x\omega + k_\perp^2)^2}{2\omega^2 k^4}B_1  \\
	&+ \frac{k_\perp^4}{4 k^4}B_2
	+ \frac{d-2}{2(d-1)}\frac{k_\perp^4}{k^4}C \, ,
\end{split}
\end{equation}
for any $d\geq 2$. In the main text $d=3$.

\section{Finite volume effects}
\label{sec:finiteV}

\subsection{The bound at finite volume}\label{sec:vevs}

Because the fireball has finite spatial extent, we should strictly work at finite volume from the start. This affects both the diagonal and off-diagonal terms in the inequality, but we will argue that it does not qualitatively change the results.

Regarding the diagonal terms, there are two sources of finite volume corrections to expectation values. The first are statistical fluctuations among the different finite volume energy eigenstates in the ETH energy window. Although these corrections are not small, they will be negative --- hence favoring violation of the ANEC --- for half of the eigenstates. The minimal eigenvalue of the off-diagonal part of the ETH matrix restricted to this negative-correction subspace will be comparable to (\ref{genmin}), because the eigenvectors of the off-diagonal matrix are randomly related to the energy eigenstates. Therefore our argument goes through. The second are positive corrections to $\varepsilon + P$ coming from, for example, spatial gradients in the temperature due to the finite volume and the vacuum state outside the fireball. These positive terms can in principle swamp out the fluctuation contribution on the right hand side of (\ref{eq_bound_manip}), even at large $L$. However, we saw in the main text that the fluctuation term becomes large at large viscosity or small sound speed. The finite size correction to the static expectation value is not expected to depend on these transport coefficients, and therefore there would seem to be no a priori reason why it should become large in tandem with the fluctuations.

Now turning to the off-diagonal terms, we assume, as discussed in the main text, that the ETH ansatz can be used for finite volume systems with size greater than the thermalization length. We would then like to relate the off-diagonal terms in the ETH ansatz to the finite volume Green's function of the energy-momentum tensor. The first complication that arises is that the Green's function in position space now depends explicitly on two distinct positions, not just on their separation. This can be written as $G_{\ocal\ocal}(t,\Delta x,x_\text{cm})$. Here $\Delta x = x_2 - x_1$ while $x_\text{cm} = (x_1 + x_2)/2$. We can Fourier transform with respect to $t$ and $\Delta x$ to obtain $G_{\ocal\ocal}(\omega, k; x_\text{cm})$. The essential new aspect of the finite volume problem is the dependence on the `center of mass' coordinate $x_\text{cm}$, due to the absence of translation invariance.

The argument in section \ref{sec:ttt} relating the off-diagonal ETH ansatz terms to the Green's function required translation invariance. This argument can only go through at finite volume if the momentum is large: $k \gg 1/L$ and $k \gg 1/(L - x_\text{cm})$. This amounts to considering pairs of points that are close relative to the scales over which finite size effects are important. If we consider a large region so that $L \gg \ell_\text{th}$, then for most points in the region, that are not too close to the boundary, $k \sim 1/\ell_\text{th}$ does the job. The ANEC involves an integral over a null ray which necessarily includes some points that are close to the boundary, but these are suppressed when $L \gg \ell_\text{th}$. Translating both the positions to the nearby center of mass position, positivity of the ETH matrix then requires
\begin{equation}\label{eq_G_finite_bound}
\frac{1}{N} \sum_{ab}\frac{G_{\ocal\ocal}(p_b-p_a;x_\text{cm})}{\Omega(p_b)} \leq |\langle \ocal(x_\text{cm})\rangle_T|^2 \, .
\end{equation}

It remains to show that $x_\text{cm}$ dependence of the Green's function does not fundamentally change the bound discussed in the main text. In particular, we need to check that the low frequency region of the integral in (\ref{eq_bound_manip}) does not get modified, as this region is responsible for the lower bound on thermalization length.

\subsection{Finite volume Green's function}

The bound \eqref{eq_G_finite_bound} involves the finite volume Green's function.
Throughout the main text we have instead used the (more convenient) infinite volume Green's function, derived in Appendix \ref{app_hyd}. At large enough wavevectors $kL \gg 1$ and frequencies $\omega \tau_L \gg 1$, where $\tau_L$ is the Thouless time\footnote{This is the time it takes for modes to cross the whole volume. For diffusive modes $\tau_L = L^2/D$, whereas for sound modes $\tau_L=L/c_s$. In general we will define $\tau_L$ to be the smallest such time scale.}, we expect both Green's functions to match. In this section we will show in fact that so long as $kL \gg 1$ the Green's function also matches at all frequencies, up to an overall nonsingular multiplicative number.

Retarded Green's functions are solutions to linear differential equations and relate expectation values of operators to sources
\begin{equation}
\langle \mathcal{O} (x,t)\rangle
	= \int dt' d^dx'\, G^R(x,x';t-t')\delta \phi(x',t')\, .
\end{equation}
Finite volume Greens functions are solutions to the same local differential equations, but are subject to different boundary conditions. These can be obtained simply by the method of images, namely by adding sources outside of the volume in order to satisfy automatically the boundary conditions. We illustrate this procedure in 1 spatial dimension $x\in V=(-L/2,L/2)$, with (for illustrative purposes) Dirichlet boundary conditions
\begin{equation}
\langle \mathcal{O} (x,t)\rangle|_{x\in \d V} = 0\, .
\end{equation}
Assuming parity invariance and focusing on parity-even operators (generalization to other cases is straightforward), the boundary conditions are satisfied if the sources are mirrored as
\begin{equation}
\delta\phi(x,t) \to
	\delta\tilde \phi(x,t) = \sum_n (-1)^n \delta\phi(nL+(-1)^nx,t)\, .
\end{equation}
The finite volume Green's function $\tilde G^R$ is therefore related to the infinite volume one as
\begin{equation}
\tilde G^R(x,x';\omega)
	= \sum_n (-1)^n G^R(nL+(-1)^nx'- x;\omega)\, .
\end{equation}
Writing $x_{\rm cm} = \frac{x'+x}{2}$ and $\Delta x = x'-x$ we have
\begin{equation}\label{eq_GG_relation}
\tilde G^R(\Delta x,x_{\rm cm};\omega)
	= \sum_{n\ \rm even}G^R(nL+\Delta x,\omega) + \sum_{n\ \rm odd} G^R(nL -2 x_{\rm cm},\omega) \,.
\end{equation}
Fourier transforming one finds after some algebra
\begin{equation}\label{eq_G_to_G}
\begin{split}
\tilde G^R(\omega,k; x_{\rm cm})
	&\equiv \int_{-(L-2|x_{\rm cm}|)}^{L-2|x_{\rm cm}|} d\Delta x\, e^{i\Delta x k}\, \tilde G^R(\Delta x, x_{\rm cm};\omega) \\
	&= \sum_{q_m = \frac{\pi m}{L}} \left[\delta_{q_m, k} + f(q_m, k, x_{\rm cm})\right]G^R(\omega, q_m)\, ,
\end{split}
\end{equation}
with 
\begin{equation}\label{eq:terms}
f(q_m, k, x_{\rm cm})
	=- \frac{2|x_{\rm cm}|}{L}\delta_{q_m,k} -(1-\delta_{q_m,k})\frac{\sin \left[2|x_{\rm cm}|(k-q_m)\right]}{L(k-q_m)}
	+ (-1)^m e^{i2q_mx_{\rm cm}} \frac{\sin 2k|x_{\rm cm}|}{Lk}\, .
\end{equation}

Equation (\ref{eq_G_to_G}) relates the finite volume Green's function $\tilde G^R$ to the infinite volume Green's function $G^R$. Firstly, note that the function $f(q_m, k, x_{\rm cm})$ is order one and nonsingular for all values of its arguments, since $|\sin a|\leq |a|$. The remaining concern is that the sum over the $q_m \neq k$ in (\ref{eq_G_to_G}) might lead to a large contribution to $\tilde G^R$ that is absent from $G^R$. However, the final two terms in (\ref{eq:terms}), that can contribute for $q_m \neq k$, are in fact small in the limit $k L \gg 1$ that we have taken, unless $|q_m - k| \lesssim 1/L$ in the second to last term. However, the scale $1/L$ is smaller than the scales appearing in the infinite volume Green's function, and therefore over this range of $q_m$ we have effectively $q_m \approx k$ in $G^R$. This means that we end up with at most a function weakly dependent on $x_\text{cm}$ multiplying the infinite volume Green's function. Therefore, in the limit $k L \gg 1$, finite size effects can alter the overall magnitude of the Green's function, but do not lead to additional singular behavior. The results in the main text will therefore survive.

\providecommand{\href}[2]{#2}\begingroup\raggedright\endgroup


\begin{thebibliography}{10}

\bibitem{Kovtun:2004de}
P.~Kovtun, D.~T. Son and A.~O. Starinets, {{Viscosity in strongly interacting
  quantum field theories from black hole physics}},
  \href{http://dx.doi.org/10.1103/PhysRevLett.94.111601}{Phys. Rev. Lett. {\bf
  94}, 111601, 2005},
  [\href{http://arxiv.org/abs/arXiv:hep-th/0405231}{{arXiv:hep-th/0405231
  [hep-th]}}].

\bibitem{Maldacena:2015waa}
J.~Maldacena, S.~H. Shenker and D.~Stanford, {{A bound on chaos}},
  \href{http://dx.doi.org/10.1007/JHEP08(2016)106}{JHEP {\bf 08}, 106, 2016},
  [\href{http://arxiv.org/abs/arXiv:1503.01409}{{arXiv:1503.01409 [hep-th]}}].

\bibitem{Cremonini:2011iq}
S.~Cremonini, {{The Shear Viscosity to Entropy Ratio: A Status Report}},
  \href{http://dx.doi.org/10.1142/S0217984911027315}{Mod. Phys. Lett. {\bf
  B25}, 1867--1888, 2011},
  [\href{http://arxiv.org/abs/arXiv:1108.0677}{{arXiv:1108.0677 [hep-th]}}].

\bibitem{Hartman:2017hhp}
T.~Hartman, S.~A. Hartnoll and R.~Mahajan, {{Upper Bound on Diffusivity}},
  \href{http://dx.doi.org/10.1103/PhysRevLett.119.141601}{Phys. Rev. Lett. {\bf
  119}, 141601, 2017},
  [\href{http://arxiv.org/abs/arXiv:1706.00019}{{arXiv:1706.00019 [hep-th]}}].

\bibitem{Baier:2007ix}
R.~Baier, P.~Romatschke, D.~T. Son, A.~O. Starinets and M.~A. Stephanov,
  {{Relativistic viscous hydrodynamics, conformal invariance, and holography}},
  \href{http://dx.doi.org/10.1088/1126-6708/2008/04/100}{JHEP {\bf 04}, 100,
  2008}, [\href{http://arxiv.org/abs/arXiv:0712.2451}{{arXiv:0712.2451
  [hep-th]}}].

\bibitem{Poland2016}
D.~Poland and D.~Simmons-Duffin, {The conformal bootstrap},
  \href{http://dx.doi.org/10.1038/nphys3761}{Nature Physics {\bf 12}, 535,
  2016}.

\bibitem{Klinkhammer:1991ki}
G.~Klinkhammer, {{Averaged energy conditions for free scalar fields in flat
  space-times}}, \href{http://dx.doi.org/10.1103/PhysRevD.43.2542}{Phys. Rev.
  {\bf D43}, 2542--2548, 1991}.

\bibitem{Faulkner:2016mzt}
T.~Faulkner, R.~G. Leigh, O.~Parrikar and H.~Wang, {{Modular Hamiltonians for
  Deformed Half-Spaces and the Averaged Null Energy Condition}},
  \href{http://dx.doi.org/10.1007/JHEP09(2016)038}{JHEP {\bf 09}, 038, 2016},
  [\href{http://arxiv.org/abs/arXiv:1605.08072}{{arXiv:1605.08072 [hep-th]}}].

\bibitem{Hartman:2016lgu}
T.~Hartman, S.~Kundu and A.~Tajdini, {{Averaged Null Energy Condition from
  Causality}}, \href{http://dx.doi.org/10.1007/JHEP07(2017)066}{JHEP {\bf 07},
  066, 2017}, [\href{http://arxiv.org/abs/arXiv:1610.05308}{{arXiv:1610.05308
  [hep-th]}}].

\bibitem{Roman:1986tp}
T.~A. Roman, {{Quantum Stress Energy Tensors and the Weak Energy Condition}},
  \href{http://dx.doi.org/10.1103/PhysRevD.33.3526}{Phys. Rev. {\bf D33},
  3526--3533, 1986}.

\bibitem{Borde:1987qr}
A.~Borde, {{Geodesic focusing, energy conditions and singularities}},
  \href{http://dx.doi.org/10.1088/0264-9381/4/2/015}{Class. Quant. Grav. {\bf
  4}, 343--356, 1987}.

\bibitem{Hofman:2008ar}
D.~M. Hofman and J.~Maldacena, {{Conformal collider physics: Energy and charge
  correlations}}, \href{http://dx.doi.org/10.1088/1126-6708/2008/05/012}{JHEP
  {\bf 05}, 012, 2008},
  [\href{http://arxiv.org/abs/arXiv:0803.1467}{{arXiv:0803.1467 [hep-th]}}].

\bibitem{Hofman:2009ug}
D.~M. Hofman, {{Higher Derivative Gravity, Causality and Positivity of Energy
  in a UV complete QFT}},
  \href{http://dx.doi.org/10.1016/j.nuclphysb.2009.08.001}{Nucl. Phys. {\bf
  B823}, 174--194, 2009},
  [\href{http://arxiv.org/abs/arXiv:0907.1625}{{arXiv:0907.1625 [hep-th]}}].

\bibitem{Chowdhury:2012km}
D.~Chowdhury, S.~Raju, S.~Sachdev, A.~Singh and P.~Strack, {{Multipoint
  correlators of conformal field theories: implications for quantum critical
  transport}}, \href{http://dx.doi.org/10.1103/PhysRevB.87.085138}{Phys. Rev.
  {\bf B87}, 085138, 2013},
  [\href{http://arxiv.org/abs/arXiv:1210.5247}{{arXiv:1210.5247
  [cond-mat.str-el]}}].

\bibitem{Cordova:2017zej}
C.~Cordova, J.~Maldacena and G.~J. Turiaci, {{Bounds on OPE Coefficients from
  Interference Effects in the Conformal Collider}},
  \href{http://dx.doi.org/10.1007/JHEP11(2017)032}{JHEP {\bf 11}, 032, 2017},
  [\href{http://arxiv.org/abs/arXiv:1710.03199}{{arXiv:1710.03199 [hep-th]}}].

\bibitem{Cordova:2017dhq}
C.~Cordova and K.~Diab, {{Universal Bounds on Operator Dimensions from the
  Average Null Energy Condition}},
  \href{http://dx.doi.org/10.1007/JHEP02(2018)131}{JHEP {\bf 02}, 131, 2018},
  [\href{http://arxiv.org/abs/arXiv:1712.01089}{{arXiv:1712.01089 [hep-th]}}].

\bibitem{PhysRevA.43.2046}
J.~M. Deutsch, {Quantum statistical mechanics in a closed system},
  \href{http://dx.doi.org/10.1103/PhysRevA.43.2046}{Phys. Rev. A {\bf 43},
  2046--2049, 1991}.

\bibitem{PhysRevE.50.888}
M.~Srednicki, {Chaos and quantum thermalization},
  \href{http://dx.doi.org/10.1103/PhysRevE.50.888}{Phys. Rev. E {\bf 50},
  888--901, 1994}.

\bibitem{rigoleth}
M.~Rigol, V.~Dunjko and M.~Olshanii, {Thermalization and its mechanism for
  generic isolated quantum systems},
  \href{http://dx.doi.org/10.1038/nature06838}{Nature {\bf 452}, 854, 2008}.

\bibitem{eth}
L.~D'Alessio, Y.~Kafri, A.~Polkovnikov and M.~Rigol, {{From quantum chaos and
  eigenstate thermalization to statistical mechanics and thermodynamics}},
  \href{http://dx.doi.org/10.1080/00018732.2016.1198134}{Adv. Phys. {\bf 65},
  239--362, 2016},
  [\href{http://arxiv.org/abs/arXiv:1509.06411}{{arXiv:1509.06411
  [cond-mat.stat-mech]}}].

\bibitem{Brehm:2018ipf}
E.~M. Brehm, D.~Das and S.~Datta, {{Probing thermality beyond the diagonal}},
  2018, [\href{http://arxiv.org/abs/arXiv:1804.07924}{{arXiv:1804.07924
  [hep-th]}}].

\bibitem{Romero-Bermudez:2018dim}
A.~Romero-Berm\'udez, P.~Sabella-Garnier and K.~Schalm, {{A Cardy formula for
  off-diagonal three-point coefficients; or, how the geometry behind the
  horizon gets disentangled}},  2018,
  [\href{http://arxiv.org/abs/arXiv:1804.08899}{{arXiv:1804.08899 [hep-th]}}].

\bibitem{Hikida:2018khg}
Y.~Hikida, Y.~Kusuki and T.~Takayanagi, {{ETH and Modular Invariance of 2D
  CFTs}},  2018,
  [\href{http://arxiv.org/abs/arXiv:1804.09658}{{arXiv:1804.09658 [hep-th]}}].

\bibitem{Son:2007vk}
D.~T. Son and A.~O. Starinets, {{Viscosity, Black Holes, and Quantum Field
  Theory}}, \href{http://dx.doi.org/10.1146/annurev.nucl.57.090506.123120}{Ann.
  Rev. Nucl. Part. Sci. {\bf 57}, 95--118, 2007},
  [\href{http://arxiv.org/abs/arXiv:0704.0240}{{arXiv:0704.0240 [hep-th]}}].

\bibitem{Kovtun:2012rj}
P.~Kovtun, {{Lectures on hydrodynamic fluctuations in relativistic theories}},
  \href{http://dx.doi.org/10.1088/1751-8113/45/47/473001}{J. Phys. {\bf A45},
  473001, 2012}, [\href{http://arxiv.org/abs/arXiv:1205.5040}{{arXiv:1205.5040
  [hep-th]}}].

\bibitem{Herzog:2007ij}
C.~P. Herzog, P.~Kovtun, S.~Sachdev and D.~T. Son, {{Quantum critical
  transport, duality, and M-theory}},
  \href{http://dx.doi.org/10.1103/PhysRevD.75.085020}{Phys. Rev. {\bf D75},
  085020, 2007},
  [\href{http://arxiv.org/abs/arXiv:hep-th/0701036}{{arXiv:hep-th/0701036
  [hep-th]}}].

\bibitem{Garrison:2015lva}
J.~R. Garrison and T.~Grover, {{Does a single eigenstate encode the full
  Hamiltonian?}},  2015,
  [\href{http://arxiv.org/abs/arXiv:1503.00729}{{arXiv:1503.00729
  [cond-mat.str-el]}}].

\bibitem{PhysRevE.97.012140}
A.~Dymarsky, N.~Lashkari and H.~Liu, {Subsystem eigenstate thermalization
  hypothesis}, \href{http://dx.doi.org/10.1103/PhysRevE.97.012140}{Phys. Rev. E
  {\bf 97}, 012140, 2018}.

\bibitem{Graham:2005cq}
N.~Graham and K.~D. Olum, {{Plate with a hole obeys the averaged null energy
  condition}}, \href{http://dx.doi.org/10.1103/PhysRevD.72.025013}{Phys. Rev.
  {\bf D72}, 025013, 2005},
  [\href{http://arxiv.org/abs/arXiv:hep-th/0506136}{{arXiv:hep-th/0506136
  [hep-th]}}].

\bibitem{Dymarsky:2017zoc}
A.~Dymarsky and H.~Liu, {{Canonical Universality}},  2017,
  [\href{http://arxiv.org/abs/arXiv:1702.07722}{{arXiv:1702.07722
  [cond-mat.stat-mech]}}].

\bibitem{Dymarsky:2018ccu}
A.~Dymarsky, {{Bound on Eigenstate Thermalization from Transport}},  2018,
  [\href{http://arxiv.org/abs/arXiv:1804.08626}{{arXiv:1804.08626
  [cond-mat.stat-mech]}}].

\bibitem{Kovtun:2011np}
P.~Kovtun, G.~D. Moore and P.~Romatschke, {{The stickiness of sound: An
  absolute lower limit on viscosity and the breakdown of second order
  relativistic hydrodynamics}},
  \href{http://dx.doi.org/10.1103/PhysRevD.84.025006}{Phys. Rev. {\bf D84},
  025006, 2011}, [\href{http://arxiv.org/abs/arXiv:1104.1586}{{arXiv:1104.1586
  [hep-ph]}}].

\bibitem{Heinz:2001xi}
U.~W. Heinz and P.~F. Kolb, {{Early thermalization at RHIC}},
  \href{http://dx.doi.org/10.1016/S0375-9474(02)00714-5}{Nucl. Phys. {\bf
  A702}, 269--280, 2002},
  [\href{http://arxiv.org/abs/arXiv:hep-ph/0111075}{{arXiv:hep-ph/0111075
  [hep-ph]}}].

\bibitem{SONG2013114c}
H.~Song, {{QGP viscosity at RHIC and the LHC - a 2012 status report}},
  \href{http://dx.doi.org/10.1016/j.nuclphysa.2013.01.052}{Nucl. Phys. {\bf
  A904-905}, 114c--121c, 2013},
  [\href{http://arxiv.org/abs/arXiv:1210.5778}{{arXiv:1210.5778 [nucl-th]}}].

\bibitem{Song:2011hk}
H.~Song, S.~A. Bass, U.~Heinz, T.~Hirano and C.~Shen, {{Hadron spectra and
  elliptic flow for 200 A GeV Au+Au collisions from viscous hydrodynamics
  coupled to a Boltzmann cascade}},
  \href{http://dx.doi.org/10.1103/PhysRevC.83.054910,
  10.1103/PhysRevC.86.059903}{Phys. Rev. {\bf C83}, 054910, 2011},
  [\href{http://arxiv.org/abs/arXiv:1101.4638}{{arXiv:1101.4638 [nucl-th]}}].

\bibitem{Cheng:2007jq}
M.~Cheng et~al., {{The QCD equation of state with almost physical quark
  masses}}, \href{http://dx.doi.org/10.1103/PhysRevD.77.014511}{Phys. Rev. {\bf
  D77}, 014511, 2008},
  [\href{http://arxiv.org/abs/arXiv:0710.0354}{{arXiv:0710.0354 [hep-lat]}}].

\bibitem{Borsanyi:2010cj}
S.~Borsanyi, G.~Endrodi, Z.~Fodor, A.~Jakovac, S.~D. Katz, S.~Krieg, C.~Ratti
  and K.~K. Szabo, {{The QCD equation of state with dynamical quarks}},
  \href{http://dx.doi.org/10.1007/JHEP11(2010)077}{JHEP {\bf 11}, 077, 2010},
  [\href{http://arxiv.org/abs/arXiv:1007.2580}{{arXiv:1007.2580 [hep-lat]}}].

\bibitem{Meyer:2007dy}
H.~B. Meyer, {{A Calculation of the bulk viscosity in SU(3) gluodynamics}},
  \href{http://dx.doi.org/10.1103/PhysRevLett.100.162001}{Phys. Rev. Lett. {\bf
  100}, 162001, 2008},
  [\href{http://arxiv.org/abs/arXiv:0710.3717}{{arXiv:0710.3717 [hep-lat]}}].

\bibitem{Bousso:2015mna}
R.~Bousso, Z.~Fisher, S.~Leichenauer and A.~C. Wall, {{Quantum focusing
  conjecture}}, \href{http://dx.doi.org/10.1103/PhysRevD.93.064044}{Phys. Rev.
  {\bf D93}, 064044, 2016},
  [\href{http://arxiv.org/abs/arXiv:1506.02669}{{arXiv:1506.02669 [hep-th]}}].

\bibitem{Bousso:2015wca}
R.~Bousso, Z.~Fisher, J.~Koeller, S.~Leichenauer and A.~C. Wall, {{Proof of the
  Quantum Null Energy Condition}},
  \href{http://dx.doi.org/10.1103/PhysRevD.93.024017}{Phys. Rev. {\bf D93},
  024017, 2016},
  [\href{http://arxiv.org/abs/arXiv:1509.02542}{{arXiv:1509.02542 [hep-th]}}].

\bibitem{Balakrishnan:2017bjg}
S.~Balakrishnan, T.~Faulkner, Z.~U. Khandker and H.~Wang, {{A General Proof of
  the Quantum Null Energy Condition}},  2017,
  [\href{http://arxiv.org/abs/arXiv:1706.09432}{{arXiv:1706.09432 [hep-th]}}].

\bibitem{Fitzpatrick:2014vua}
A.~L. Fitzpatrick, J.~Kaplan and M.~T. Walters, {{Universality of Long-Distance
  AdS Physics from the CFT Bootstrap}},
  \href{http://dx.doi.org/10.1007/JHEP08(2014)145}{JHEP {\bf 08}, 145, 2014},
  [\href{http://arxiv.org/abs/arXiv:1403.6829}{{arXiv:1403.6829 [hep-th]}}].

\bibitem{Asplund:2014coa}
C.~T. Asplund, A.~Bernamonti, F.~Galli and T.~Hartman, {{Holographic
  Entanglement Entropy from 2d CFT: Heavy States and Local Quenches}},
  \href{http://dx.doi.org/10.1007/JHEP02(2015)171}{JHEP {\bf 02}, 171, 2015},
  [\href{http://arxiv.org/abs/arXiv:1410.1392}{{arXiv:1410.1392 [hep-th]}}].

\bibitem{Turiaci:2016cvo}
G.~Turiaci and H.~Verlinde, {{On CFT and Quantum Chaos}},
  \href{http://dx.doi.org/10.1007/JHEP12(2016)110}{JHEP {\bf 12}, 110, 2016},
  [\href{http://arxiv.org/abs/arXiv:1603.03020}{{arXiv:1603.03020 [hep-th]}}].

\bibitem{Faulkner:2017hll}
T.~Faulkner and H.~Wang, {{Probing beyond ETH at large $c$}},  2017,
  [\href{http://arxiv.org/abs/arXiv:1712.03464}{{arXiv:1712.03464 [hep-th]}}].

\bibitem{ElShowk:2012ht}
S.~El-Showk, M.~F. Paulos, D.~Poland, S.~Rychkov, D.~Simmons-Duffin and
  A.~Vichi, {{Solving the 3D Ising Model with the Conformal Bootstrap}},
  \href{http://dx.doi.org/10.1103/PhysRevD.86.025022}{Phys. Rev. {\bf D86},
  025022, 2012}, [\href{http://arxiv.org/abs/arXiv:1203.6064}{{arXiv:1203.6064
  [hep-th]}}].

\bibitem{Dymarsky:2017yzx}
A.~Dymarsky, F.~Kos, P.~Kravchuk, D.~Poland and D.~Simmons-Duffin, {{The 3d
  Stress-Tensor Bootstrap}},
  \href{http://dx.doi.org/10.1007/JHEP02(2018)164}{JHEP {\bf 02}, 164, 2018},
  [\href{http://arxiv.org/abs/arXiv:1708.05718}{{arXiv:1708.05718 [hep-th]}}].

\bibitem{Iliesiu:2018fao}
L.~Iliesiu, M.~Kologlu, R.~Mahajan, E.~Perlmutter and D.~Simmons-Duffin, {{The
  Conformal Bootstrap at Finite Temperature}},  2018,
  [\href{http://arxiv.org/abs/arXiv:1802.10266}{{arXiv:1802.10266 [hep-th]}}].

\bibitem{Hartnoll:2014lpa}
S.~A. Hartnoll, {{Theory of universal incoherent metallic transport}},
  \href{http://dx.doi.org/10.1038/nphys3174}{Nature Phys. {\bf 11}, 54, 2015},
  [\href{http://arxiv.org/abs/arXiv:1405.3651}{{arXiv:1405.3651
  [cond-mat.str-el]}}].

\end{thebibliography}
\end{document}